\documentclass[a4paper,fleqn,usenatbib]{mnras}

\usepackage[T1]{fontenc}
\usepackage{ae,aecompl}

\usepackage{graphicx}	
\usepackage{amsmath}	
\usepackage{amssymb}	

\newcommand{\II}{\,{\sevensize II}}
\newcommand{\III}{\,{\sevensize III}}

\title[SITELLE at the CFHT]{SITELLE: An Imaging Fourier Transform Spectrometer for the Canada-France-Hawaii Telescope}

\author[L. Drissen et al.]
{Laurent Drissen$^{1,2,3,4}$,\thanks{E-mail: ldrissen@phy.ulaval.ca}
Thomas Martin$^{1,2}$,
Laurie Rousseau-Nepton$^{3,4}$,
\newauthor Carmelle Robert$^{1,2,3,4}$, R. Pierre Martin$^3$,
\newauthor  Marc Baril$^{4}$, Simon Prunet$^{4,5}$, 
\newauthor Gilles Joncas$^{1,2}$, Simon Thibault$^{1,2}$, Denis Brousseau$^{1,2}$, 
\newauthor  Julie Mandar$^{6}$, Fr\'ed\'eric Grandmont$^{6}$,
\newauthor  Howard Yee$^7$, Luc Simard$^8$\\
$^{1}$D\'epartement de physique, de g\'enie physique et d'optique, Universit\'e Laval, Qu\'ebec (QC), G1V 0A6, Canada\\
$^{2}$Centre de recherche en astrophysique du Qu\'ebec\\
$^{3}$Department of Physics and Astronomy, University of Hawai'i at Hilo, Hilo, HI 96720, USA\\
$^{4}$Canada-France-Hawaii Telescope, 65-1238 Mamalahoa Hwy, Kamuela, Hawaii 96743, USA\\
$^{5}$CNRS and UPMC, UMR 7095, Institut d'Astrophysique de Paris, F-75014 Paris, France\\
$^{6}$ABB Inc., 3400 Rue Pierre-Ardouin, Qu\'ebec, QC G1P 0B2, Canada\\
$^{7}$Department of Astronomy and Astrophysics, University of Toronto, Toronto, ON M5S 3H4, Canada\\
$^{8}$National Research Council Canada, 5071 West Saanich Roard, Victoria, BC, V9E 2E7, Canada\\
}
\date{Accepted XXX. Received YYY; in original form ZZZ}

\pubyear{2016}

\begin{document}
\label{firstpage}
\pagerange{\pageref{firstpage}--\pageref{lastpage}}
\maketitle

\begin{abstract}
We present an overview of SITELLE, an Imaging Fourier Transform Spectrometer (iFTS) available at the 3.6-meter Canada-France-Hawaii Telescope. SITELLE  is a Michelson-type interferometer able to reconstruct the spectrum of every light source within its 11$'$ field of view in filter-selected bands of the visible (350 to 900\,nm). The spectral resolution can be adjusted up to R\,=\,10\,000 and the spatial resolution is seeing-limited and sampled at 0.32$''$ per pixel. We describe the design of the instrument as well as the data reduction and analysis process.  To illustrate SITELLE's capabilities, we present some of the data obtained during and since the August 2015 commissioning run. In particular, we demonstrate its ability to separate the components of the [OII] $\lambda\lambda$ 3726,29 doublet in Orion and to reach R = 9500 around H$\alpha$; to detect diffuse emission at a level of 4 $\times$\,10$^{-17}$\,erg~cm$^{-2}$\,s$^{-1}$\,arcsec$^{-2}$; to obtain integrated spectra of stellar absorption lines in galaxies despite the well-known multiplex disadvantage of the iFTS; and to detect emission-line galaxies at different redshifts.
\end{abstract}




\section{Introduction}

Fourier transform spectrometers (FTS) have a long history of astronomical applications, both ground- and space-based, ranging from the study of individual stars and galactic nuclei \citep{1977ApJS...34..101P,1977ApJS...35..397P}, planetary atmospheres \citep{1988Sci...240.1767O}, the cosmic microwave background \citep{1990PhRvL..65..537G,1990ApJ...354L..37M} and, more recently, galactic nebulae and star-forming regions \citep{2016MNRAS.458.2150M}.

Adding an imaging capability to the original single aperture FTS was an obvious step forward, as was the advent of the first dispersive integral field spectrographs. One of the first of these imaging FTS (iFTS), BEAR \citep{1996ASPC..102..232S,1997A&A...321..907C,2000ASPC..195..185M}, attached to the Canada-France-Hawaii Telescope (CFHT), was capable of reaching R\,$\simeq$\,10$^4$  for spectra in the  1\,-\,5\,$\mu$m range of extended objects with a 24$''$ circular field of view (FOV) and a sampling of 0.25$''$/pixel. A decade later, a prototype iFTS working in the visible band was designed and built at the Lawrence Livermore National Laboratory (LLNL), and was successfully used to gather spatially resolved spectra of bright targets \citep{2002ASPC..280..139W}.The development of this instrument was a major step forward to demonstrate the ability of an imaging FTS to acquire hyperspectral images in the visible band.

The relative merits of different instruments for 3D spectroscopy, including FTS, have been discussed by \citet{1995ASPC...71..328S} and \citet{2000ASPC..195...58B}. A more recent and detailed analysis of the advantages and disadvantages of the iFTS solution compared to other integral field spectrographs was presented by 
\citet{2013ExA....35..527M}. 

Motivated by the NGST Science and Technology Exposition conference in which several teams presented designs of iFTS for what would become the James Webb Space Telescope \citep{2000ASPC..207..240G,2000ASPC..207..276M,2000ASPC..207..303P}, and encouraged by the success of the LLNL group, our team, working with the Qu\'ebec City-based high-tech company Bomem (later, included in ABB), has developed a wide-field 
($12' \times 12'$) iFTS, SpIOMM  \citep{2003SPIE.4842..392G,2006SPIE.6269E..49B,2008SPIE.7014E..7JB}, attached to the 1.6-m telescope of the Observatoire du 
Mont-M\'egantic and used it to study Galactic nebulae and nearby galaxies in selected passbands of the visible range \citep{2008SPIE.7014E..7KD,2010AJ....139.2083C,2012MNRAS.420.2280L,2015MNRAS.448.1584L}.

\begin{figure}
	\includegraphics[width=\columnwidth]{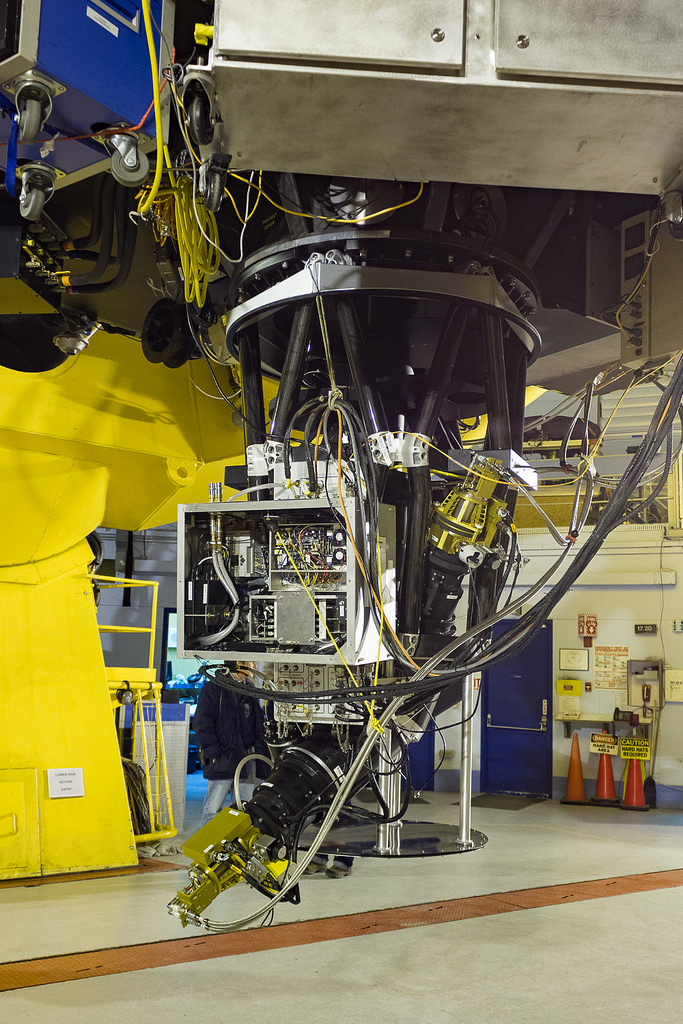}
    \caption{ SITELLE at the Cassegrain focus of the Canada-France-Hawaii Telescope (Courtesy Tom Benedict). 
    The two CCD enclosures are in golden color.}
    \label{fig:Sitelle}
\end{figure}

We present in this paper the characteristics of SITELLE, an improved version of SpIOMM working in the visible band (350\,-\,900\,nm), designed for the CFHT. As a guest instrument, SITELLE is being used on a regular basis since January 2016 (see Figure~\ref{fig:Sitelle}).
We also highlight its capabilities by showing some technical and science results during and since its commissioning and science verification observing runs. 

Most of the design and construction work has been performed by ABB Analytical, a Qu\'ebec-based company specialized
in Fourier transform spectrometers and optical sensors. Science lead, optical design, and its integration were done at
Universit\'e Laval, the mechanical design and fabrication of the input and output optics at Universit\'e de Montr\'eal, while
CFHT took the responsibility of the detectors' enclosure and cooling system.

\section{Science Drivers and Design Constraints}
The number of science cases for a wide-field imaging spectrograph is very large. We have however
selected three research areas taking advantage of the iFTS capabilities to drive the instrument design: the study
of the physical characteristics (temperature, density, abundances, and kinematics) of nebulae surrounding evolved stars, supernova remnants,
and the diffuse interstellar gas in the Milky Way; the structure and abundances of large numbers of H\II\ regions in nearby galaxies; and the detection and characterization of emission-line galaxies in nearby clusters. All these projects aim at emission-line targets. This is because the well-known multiplex disadvantage of the FTS \citep{2013ExA....35..527M} makes it much less competitive than a dispersive IFS of similar field of view for the study of continuum and absorption-lines objects, whereas extended objects with emission lines spectra are perfectly suited for this type of instrument. As will be shown below, SITELLE is capable of measuring absorption features, but is not optimized for it.

These programs have defined the following technical requirements for SITELLE which have shaped the
design of the instrument (as summarized in Table\,\ref{tab:instrument}):

\begin{table*}
	\centering
	\caption{SITELLE Characteristics}
	\label{tab:instrument}
	\begin{tabular}{ll}
		\hline
		Field of View&  $11'  \times 11'$ \\
		Pixel size&  0.32$''$ \\
		Detectors&  2\,$\times$\,2048$^2$ Deep depletion e2v \\
		Readout Noise&  4.5\,e- \\
		Readout Time&  3.8\,s \\	
		Spectral resolution&1\,-\,10\,000 \\
		Wavelength Range&350\,-\,900 nm\\
		Declination limits & $-$38$^{\circ}$ to +67$^{\circ}$ \\			
		\hline
	\end{tabular}
\end{table*}

{\it Wavelength Range} - The need to detect the [O\II]\,$\lambda 3727$ doublet defines the short wavelength requirement.
It is used to measure the oxygen abundance in ionized nebulae and the ratio of its two components is an
excellent indicator of the electron density in H\II\ regions. Many factors conspire to make this line a
real challenge for an iFTS, and in particular the stringent constraints it imposes on the quality of the
optical surfaces within the interferometer (mirrors and beamsplitter) as well as the precision of the step-scan
and servo mechanisms to which the modulation efficiency of the interferometer - its capability to maximize the fringe contrast and therefore extract spectral information from the interferograms - is particularly sensitive at short wavelengths. The long wavelength limit was originally defined by the desire to observe the [S\II]\,$\lambda\lambda$6717,6731 doublet in galaxies at a redshift of $\sim$\,0.017. Although pushing this limit to the red in order to reach some of the OH windows below 900\,nm for cosmology applications posed no challenge to the instrument design itself, it imposed a stringent limit on the amplitude of the CCD fringes. In fact, fringes are even sometimes seen at wavelengths close to H$\alpha$; this effect can be corrected for the purpose of pure imagery \citep{2012PASP..124..263H}, but becomes a severe constraint when interferograms are concerned.

{\it Spectral Resolution} - The minimum resolution required for the analysis of the ionized nebula in the Milky Way
and other galaxies is set by the necessity to separate the [S\II]\,$\lambda\lambda 6717,6731$ doublet, the H$\alpha$ line from
its [N\II]\,$\lambda\lambda$6548,6584 adjacent lines, and H$\gamma$ from [O\III]\,$\lambda 4363$. This implies a minimum
value of R\,=\,1000 over the entire wavelength range. However, the kinematics studies of H\II\ regions and nearby galaxies, as well as the possibility to determine electron densities with the (often very strong) [O\II] doublet ratio imposed a 
more stringent minimal requirement of R\,$\simeq$\,3000. As explained below, the use of filters to isolate a group of lines allows to increase the spectral resolution for a given number of mirror steps.

{\it Field of View and Pixel Size} -  For the study of extended H\II\ regions in the Milky Way, nearby galaxies, and cluster of galaxies,
 reasonable amounts of observing time require a FOV larger than $\sim$\,10$'$, which set the constraints for
SITELLE. A larger FOV is always welcome but leads to instrumental design challenges that are not linear with the field
size beyond this point. The specified value appears as a sweet spot considering the available budget.
In terms of spatial sampling and image quality, the instrument had to take advantage of the excellent image quality of
the CFHT.

{\it Overheads} - Obtaining a datacube with an iFTS typically requires a few hundred CCD readouts and Michelson interferometer mirror displacements. In order to minimize overheads, the CCD readout time was required to be comparable to the time it takes to move the mirror and stabilize the interferometer, the two being concurrent, and had to be less than 5\,seconds, without a significant increase in readout noise.

{\it Filters} - Filters selecting bandpasses of interest across the visible range must be used to decrease the photon noise and increase the spectral resolution for a given number of mirror steps; more details will be provided below.

\section{The instrument}

\subsection{General Design}
An astronomical iFTS is basically a Michelson interferometer inserted into the collimated beam of an astronomical camera system equipped with two detectors. Spectra of every source of light in the FOV (11$'$\,$\times$\,11$'$ in the case of SITELLE) are reconstructed from a series of images obtained by moving one of the two mirrors of the interferometer, producing an interferometric cube which is then Fourier transformed to produce a spectral datacube. All wavelengths within the range set by the filter used are simultaneously transmitted to either one or both of the interferometer outputs in which the array detectors sit. By moving one of its two mirrors, thereby changing the optical path difference (OPD), the interferometer thus configured therefore modulates the scene intensity between the two outputs instead of spectrally filtering it (see Section~\ref{sec:deep} for an illustration). This configuration results in a large light gathering power since no light is lost except through items common to any optical design: substrate transmission, coatings efficiency, and quantum efficiency of the detectors. All photons from the source can hence be recorded at each exposure provided that both complementary outputs of the interferometer are recorded. This requires a modification to the standard Michelson configuration in which half the light goes back to the source: the incoming light enters the interferometer at an angle allowing the two output beams to be physically separated. A CCD detector is then attached to each of the two output optic ports collecting the light from the interferometer.

\begin{figure}
	\includegraphics[width=\columnwidth]{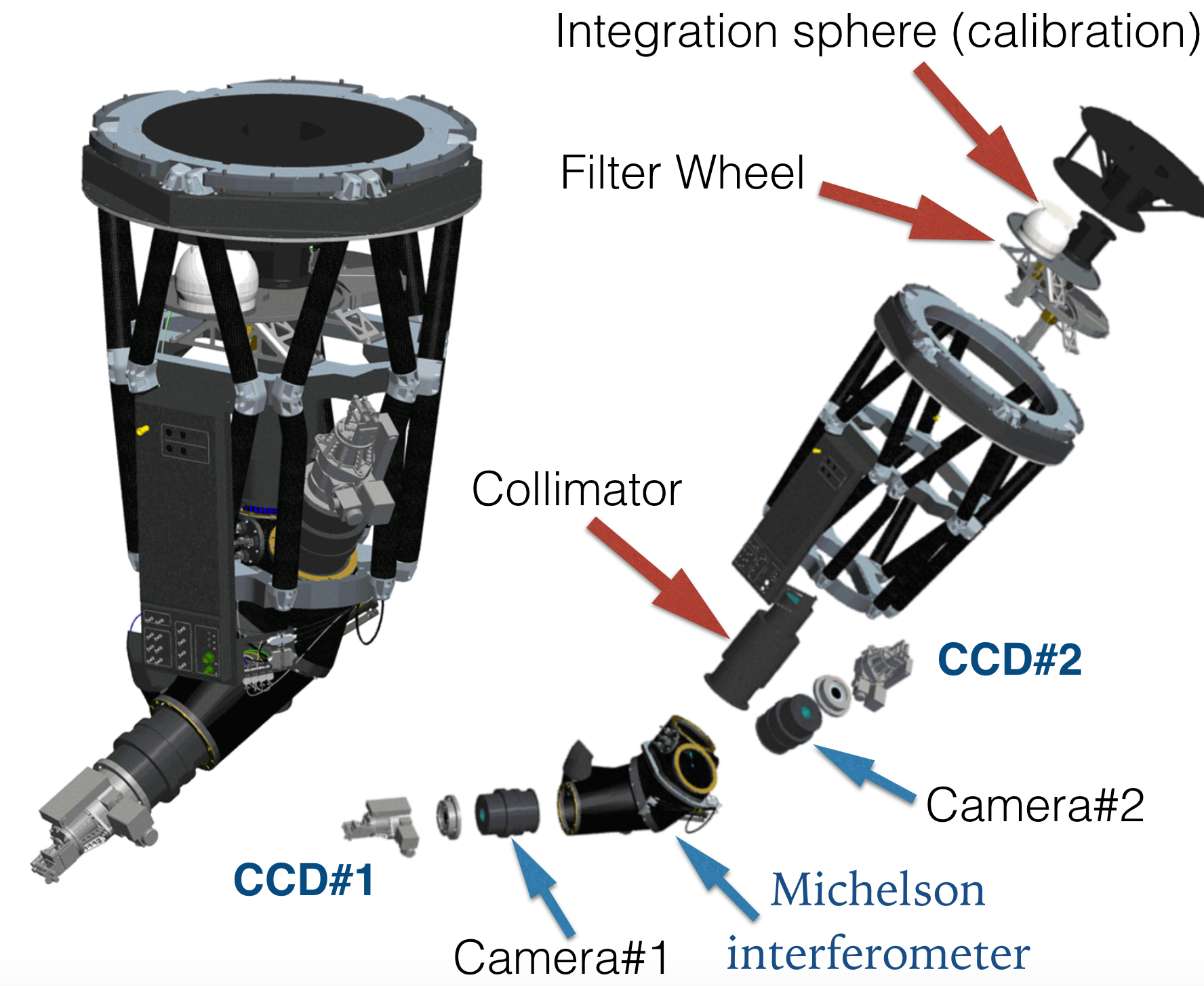}
    \caption{SITELLE exploded view}
    \label{fig:exploded}
\end{figure}

\begin{figure}
	\includegraphics[width=\columnwidth]{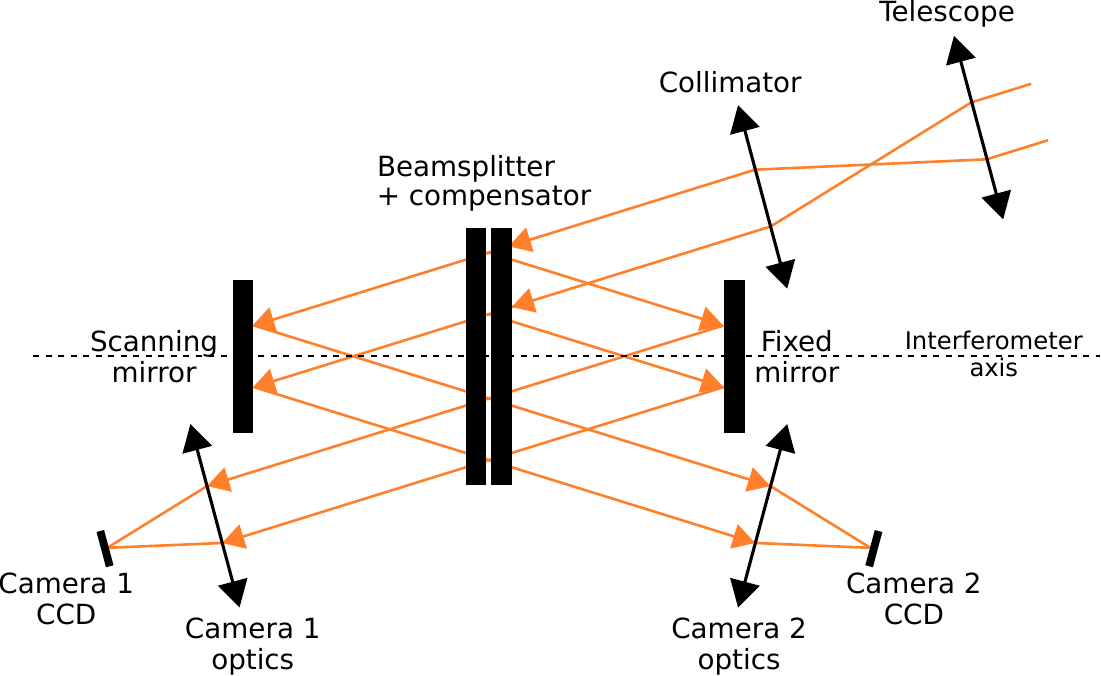}
    \caption{Simplified 2D representation of the optical arrangement of SITELLE. }
    \label{fig:Interfero}
\end{figure}

In addition to its mechanical structure and electronics, SITELLE is therefore composed of (see Figure~\ref{fig:exploded}):
\begin{itemize}
\item A filter wheel to select the appropriate bandpasses;
\item A collimator;
\item The Michelson interferometer  which consists of:

\begingroup{
\leftskip=1truecm
\parindent=-0.28truecm

- A beamsplitter/compensator used to separate the incoming beam into two equal parts (see Figure~\ref{fig:Interfero});

- Two mirrors on which the halves of the original beam are reflected back;

- A scanning mechanism to adjust the position and orientation of the scanning mirror (the other mirror is fixed);

- A metrology system (IR laser and detector) to monitor the mirror alignment and position;

}\endgroup

\item Two output camera optics;
\item Two CCD detectors;
\item An integration sphere for calibrations.
\end{itemize}

Some details on the final design of SITELLE and its rationale have been presented by \citet{2012SPIE.8446E..0UG} and \citet{2014AdAst2014E...9D}.

\subsubsection{The Interferometer}
The core constituent of SITELLE is an off-axis Michelson interferometer, whose design is largely driven by the desire to 
obtain high efficiency at near UV wavelengths over a wide FOV. A very small number of interferometers  have operated in the UV regime in the literature because the modulation efficiency (which dictates
the instrument's ability to extract spectral information from the source) shows an exponential-like decline in performance towards shorter wavelengths. Many factors affect the modulation efficiency, the most important being wavefront errors, tilt or shear between the two recombining beams, as well as optical path difference jitters during an exposure.  Excellent quality of the reflecting surfaces as well as an adequate choice of interferometer configuration are key ingredients to success.

Because they are inherently tilt-error free,  cube corner retro-reflector-based interferometers are by far the most common FTS architecture nowadays.  Cube corners of acceptable quality for infrared interferometry down to 1\,$\mu$m are found
commercially in sizes up to 5\,cm, but SITELLE's requirements on the FOV and wavelength range would require cubes 
with four times better alignment and reflected wavefront errors in size exceeding 20\,cm of clear aperture. This option was considered too risky, since inability to achieve the requirements would result in irrecoverable performance losses at short wavelengths; we therefore
chose the more classical flat mirror Michelson design. 

However, as mentioned before, in a standard plane mirror Michelson interferometer, half
the light goes to one output and the other half is retro-reflected towards the target. This is clearly not desired in
ground-based astronomy since losing half the flux from the sources translates into a significant loss of telescope 
efficiency and also prevents to fully characterize and correct for fluctuation of the sky transparency during the scan of the interferograms.
Indeed, the Fourier transform technique used to recover the spectra makes no distinction  between true interferences and undesired time-dependent source fluctuations. Accessing the second output port is therefore crucial to compensate for the global 
source intensity variations before performing the Fourier transform. 
Nevertheless, this compensation does not include any spectral content variation and remains a source of uncertainties, 
marginal but inherent to the observation from a ground-based facility.

To overcome this standard Michelson configuration problem, a solution already implemented in SpIOMM was reused. It consists of
entering the interferometer at a given angle such that the coincident output is angularly separated from the input. The
angle, 15$^\circ$, is made just large enough to locate the collimator lens and the camera lens barrels side-by-side as depicted in the
simplified diagram shown in Figure~\ref{fig:Interfero}. The main effect of this off-axis design is to offset the center of the interference fringe pattern (the ``bull's eye'') from the center of the images by 15.5$^\circ$: given the characteristics of the CCD, the southern part of the field is located 11.8$^{\circ}$ from the bull's eye and the northern part, 19.6$^{\circ}$. Figure~1 in \citet{2018MNRAS.473.4130M} shows the offset angle map between the center of the fringe pattern and the position on the CCD.

In addition to using simple optical components (plane mirrors) readily available commercially, 
this approach removes two reflecting surfaces from the cube-corner design, which helps to further reduce the errors between recombining wavefronts and increase throughput. Its drawback is that a very stringent dynamic alignment system must be implemented to correct for the numerous contributors to the overall tilt of the system.

\subsubsection{Scan Mechanism, Metrology, and Dynamic Alignement}

In most commercial FTS, designed to observe very bright sources, the interferometer's mirror is moved at a regular and well monitored 
servoed speed. The weak signal 
from astronomical sources combined with the relatively slow readout rate from our detector (3.8\,s) rather suggest a very slow scanning or step 
scanning of the interferogram. Step scanning is favored as it allows using the undersampling technique efficiently which consists 
in skipping interferogram points and retrieving the spectrum over a restricted spectral range defined by an optical filter placed in the 
optical path. Step scanning allows to choose arbitrary sampling intervals and to move rapidly between OPD positions where long 
exposure can be made. The OPD interval between steps is selected to retrieve a spectral range slightly larger than that of the filter. The interferogram cube is thus obtained through the acquisition of a series of short exposures (typically $\sim$\,10\,s to $\sim$\,2\,min) with the CCDs. Between each step, the scanning mirror in the interferometer is displaced by a very short distance (between  $\sim$\,0.5 and 5\,$\mu$m), depending on the spectral resolution and spectral band  (see Table~\ref{tab:filters_table}).

\begin{table*}
	\centering
	\caption{Existing Filters for SITELLE}
	\label{tab:filters_table}
	\begin{tabular}{cccccl}
		\hline
		Filter& T > 90\% &  Folding & Step size &N$_{1000}^*$&Goals\\
		 & range (nm) & order & (nm) & & \\
		\hline
		SN1 & 363 - 386 & 8 & 1647 &168 & [O\II]\,$\lambda$3727 < 10\,000\,km\,s$^{-1}$ \\
		C1 & 389 - 484 & 2 & 570 & 502 & H$\delta$, H$\gamma$, [O\III]\,$\lambda$4363, \\
		 & & &  & & He\II\,$\lambda$4686, Ca H\&K,\\
		 & & & & & 4000\,\AA\ break, \\
		 & & & & & [O\II]\,$\lambda$3727 z\,=\,0.046\,-\,0.29\\
		SN2 & 482 - 513 & 6 & 1680 & 216 & H$\beta$, [O\III]\,$\lambda$5007 < 7500\,km\,s$^{-1}$\\
		C3 & 511 - 556 & 6 & 1778 & 215 & Mg, Fe absorption\\
		& & & & & [Cl\III]\,$\lambda\lambda$5517,5537 \\
		C2 & 559 - 625 & 5 & 1680 & 251 & [N\II]\,$\lambda$5755, He\,{\sevensize I}\,$\lambda$5876,\\
		& & & & & [O\II]\,$\lambda$3727 z\,=\,0.50\,-\,0.68 \\
		SN3 & 647 - 685 & 8 & 2943 & 168 & H$\alpha$, [N\II]\,$\lambda\lambda$6548,6584,\\
		& & & & & [S\II]\,$\lambda\lambda$6717,6731 < 5400\,km\,s$^{-1}$ \\
		C4 & 796 - 826 & 12 & 5270& 116 & H$\alpha$ z\,$\simeq$\,0.25\\
		\hline
$^*$ Number of steps to reach R\,=\,1000
	\end{tabular}
\end{table*}

The scan mechanism is a custom-designed architecture based on ABB's heritage in flex-blade actuated frictionless scan
mechanism. The blades arrangement surrounding the mechanism produces a purely translational movement and reduces
tilt sensitivity to gravity orientation changes with a first tilt mode at 540\,Hz. It is actuated by piezoelectric actuators. 
A {\it Physik Instrumente} Nexline actuator
is used for the coarse displacement and stacked piezos are used for fine tuning of the OPD and the mirror alignment. The absence of
static friction combined with the piezos allows for sub-nm OPD corrections. The high axial rigidity produced by the 
actuator and the piezos  also reduces sensitivity to operational vibrations by ensuring that the scanning mirror 
follows the rest of the interferometer movement, if present.

The metrology uses a 1550\,nm high-stability laser. The CCD detector is not sensitive to the laser wavelength, which
allows to constantly monitor and correct the OPD and mirror alignment during the exposure of the detectors. A multi-beam pattern
surrounding the science beam allows to easily retrieve both mirror position and angle with a precision better than
1/1000th of a laser fringe. The metrology fringe signals are digitized by an ADC and processed using an ABB
proprietary method based on quadrature fringe signal which provides continuous absolute OPD information throughout
the 1 cm scanning range at a frequency of 10\,kHz. The metrology servo does not require such a high sampling rate to
operate, but the feedback is meant to be very rapid to ensure proper safety margin between successive readings in order to avoid
confusion between adjacent fringes under rapid perturbations.

An initialization process combining broadband sources to the metrology lasers allows to retrieve absolute OPD 
down to nm precision through power cycling of the entire system. The whole metrology system makes  it  entirely  possible to  start  a  
cube  on  one  night  and  finish  it  a  few  nights  later.

\subsection{Optical Design}

Details about SITELLE's optical design, assembly, and testing are provided by \citet{2014SPIE.9147E..3ZB}, so we only present here a brief overview. SITELLE's optics consists of two groups: a collimator with a focal length of 711.3\,mm, composed of three lenses and two identical cameras with a focal length of 236.3\,mm, which contain six lenses each. The design allows for an unvignetted circular FOV of 5.5$'$  radius and a full FOV of $11' \times 11'$ (0.32$''$ per pixel) with a maximum of 15\% vignetting and an optical distortion of $\sim$\,2\% at the corners of the field. Finding suitable anti-reflection coatings which did not significantly attenuate the light was a challenge considering the large number of optical surfaces (18) within the optical path. To maintain a light throughput as flat as possible through the complete wavelength range, and more specially at the science wavelength of the [O\II] doublet, we asked the coating manufacturers to shift the coating curves by 120\,nm to the blue side of the spectrum for four of the lenses. 
Optical tests in the lab have demonstrated that the image quality values were in good agreement with the nominal ones derived from the optical design and were all below the 0.8$''$ FWHM requirement of SITELLE. However, images obtained at the telescope show a significant degradation of the otical quality at the edge of the field, in particular in the upper section. More on this is presented in Section~5.3.

\subsection{Detectors}
SITELLE is equiped with two 15\,$\mu$m 2k\,$\times$\,2k pixels e2v deep-depletion CCDs with very flat quantum efficiency above 90\% between 
400 and 780\,nm, declining on both sides to 50\% at 350 and 920\,nm. Read noise is 4.5\,e-, and the readout time is 3.8\,s from four outputs. 
Each camera is cooled using a PolyCold PCC cold-head charged with PT-14 gas to allow lower temperature operation of the activated carbon getter 
(85-93\,K typical).  The cooling system introduces a significant source of vibration, which is transmitted to the interferometer; but, as shown by \citet{2016SPIE.9908E..29B}, the metrology and piezo largely compensate for it, so the vibrations do not introduce significant noise into the spectra. 

\subsection{Filters}
SITELLE could in principle be used without any filter and obtain spectra covering the entire visible band. However, the use of filters is the norm for three reasons. First, the optics are not achromatic over the entire visible range; a filterless use would therefore degrade the image quality, although this is a rather minor point. Second, since the main disadvantage of all FTS is the distributed photon noise, using a filter strongly reduces the background, from the night sky as well as from the underlying continuum from the object itself. Finally, since the spectral resolution depends on the number of mirror steps, selecting a wavelength range including important spectral features allows to limit the number of mirror steps, and hence minimize overheads for a given spectral resolution. Spectral folding is then used to lower the number of steps while still achieving the desired spectral resolution: the mirror steps are then much larger than they would if no filter were used while still sampling the interferograms at the Nyquist frequency. This imposes stringent constraints on the out-of-band transmission in all filters. 

SITELLE is equipped with a filter wheel with six positions, five of them being reserved for filters, one of which has to be occupied by a filter allowing transmission of the calibration laser light. Characteristics of the filters are presented in Table~\ref{tab:filters_table}. All filters have been provided by the Phoenix-based Custom Scientific; they have very flat transmission curves with T$_{max} > 95\%$ and sharp edges. Most of the 360\,-\,825\,nm range is covered with the current filter set, the only significant gaps below 650\,nm are caused by the desire to avoid the bright night-sky [O\,{\sevensize I}]\,$\lambda$5577 and 6300 emission lines. The SN filter series is intended to determine the properties of H\II\ regions of the Milky Way and nearby galaxies using strong emission lines \citep{2002ApJS..142...35K}, while the C series aims at fainter lines in H\II\ regions, absorption lines in nearby galaxies, and the detection of emission line galaxies in low-redshift clusters. 

\subsection{Modulation Efficiency and Global Throughput}
\begin{figure}
	\includegraphics[width=\columnwidth]{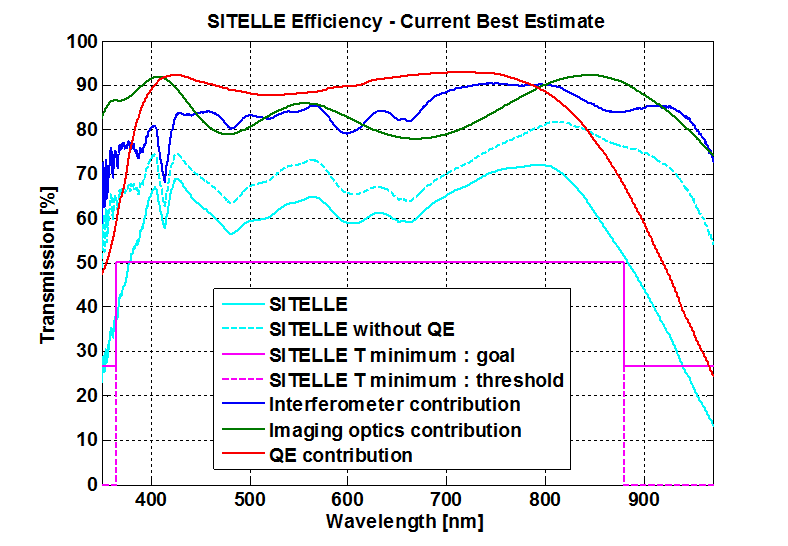}
    \caption{SITELLE throughput. The current best estimates for the transmission are shown for the different contributions as identified.}
    \label{fig:throughput}
\end{figure}

SITELLE's global throughput (excluding the telescope mirrors) is shown in Figure~\ref{fig:throughput}. All of its contributors are common to other imaging instruments, except the modulation efficiency (ME) of the interferometer (dark blue in Figure~\ref{fig:throughput}), which depends not only on the optical properties of the beamsplitter and the mirrors (the coating properties explain the wiggles seen), but also on the ability of the alignement system to minimize the mirrors' excursion from the desired position and orientation.

Variability of the modulation efficiency measured after the system initialization is small at the zero path difference (ZPD - where the optical path are identical in both arms of the interferometer) position 
(of the order of 3\%) whereas it can be greater when the moving mirror gets further. 
This can happen when the sampling of the interferogram requires greater mirror displacement, 
e.g. while scanning the red section of the spectrum at high-spectral resolution (R\,>\,5000).  
Measurement of the modulation efficiency has been extracted over the entire mirror displacement range using 
high-resolution laser datacubes. Within the usual operation range, its value is stable and can be measured directly using standard 
star datacubes observed in photometric conditions by comparing the standard stellar flux in a given aperture obtained from the unmodulated combination of interferograms on one hand, and from the integrated stellar spectrum on the other hand. Finally, note that ME variations at ZPD alone are responsible for photometric calibration errors. Indeed, ME variations outside ZPD have an impact on the shape of the ILS, but not on its integral value.

In principle, the ME values obtained from the laser images taken immediately before and after the datacube could be used to compute an expected ME curve for the science cube. In practice, several complications arise:

1) To avoid multiplying large movements of the mobile mirror - that resulted sometimes in metrology losses - the laser images are taken at an OPD equal to the OPD value at the start and end of the science datacube;

2) The variability of the measured ME curve as a function of OPD for different interferometer initialization make the prediction of the ME curve difficult for a particular initialization (where the ME is only known at a given OPD);

3) The ME computed on laser images is representative of the green HeNe laser frequency, and not necessarily of the other frequencies (with an expected degradation in the blue and improvement in the red). We thus rely on the optical model of the interferometer for these extrapolations.

The ME variability is considered to be the limiting factor to the spectrophotometric accuracy of SITELLE data, and is currently estimated to
be of the order of 5\% - 10\%.

\section{Observations, Data reduction, and Calibration}

\subsection{Observing Run Procedure}
SITELLE's observations are executed within the CFHT's queue system. 
Each queue consists in an optimized sequence of observation blocks (datacube or calibrations) covering the whole night accounting 
for all programs priority, observing constraints, and required calibrations \citep{2011tfa..confE..62M}. 
The queue system allows to split the observation of a datacube in two or more sections and observe them during different nights. 
Within a datacube, some images can also be re-observed right away or on another night if their quality is insufficient. 
A typical sequence of observations would include flats in all filters, standard star images accompanying each observing block, 
and observing blocks distributed to fill the night while accounting for the airmass and moon distance. 
Usual daytime calibration includes flat datacubes using LED lights projected on the internal integration sphere and a 
high-resolution He-Ne laser datacube, both observed at the zenith. 
Those are only required once every run and allows to produce the phase map and the spectral calibration respectively. 
Also, twice a year, standard star datacubes are acquired to measure the global transmission curve of the system in all filters.

\subsection{Data Reduction and Analysis}
The increasing complexity of the new astronomical instruments makes the
design of dedicated data reduction and analysis software a fundamental
component of the system efficiency. We have put enormous efforts on the
development of a fully automated data reduction pipeline, ORBS,
as well as a fitting engine for a fast and reliable
extraction of the spectral parameters, ORCS (Martin et al. 2018, in prep.).

\subsubsection{Data Reduction}

The details of the reduction process will be discussed elsewhere, but
we present here a summarized version of it:

\begin{enumerate}
\item Bias, Flat-field, Images Alignment, and Cosmic Rays~-~The interferomeric images are corrected for the electronic bias and the flat-field curvature. Images are then aligned to compensate for guiding errors.
 
Because any cosmic rays falling on a pixel in an image will affect the whole spectrum calculated for this pixel, 
it is important to take into account these events, which are numerous considering the high rate of the muon flux at the Mauna Kea altitude 
($\sim$\,2\,cm$^{-2}$\,min$^{-1}$;  \citealt{2002ExA....14...45G}). 
The CCD readout time being 3.8~s, it would be time-consuming to record more than one image at each OPD position in order 
to reduce the number of cosmic rays. We thus have developed a simple but efficient algorithm to detect and correct cosmic rays. 
This algorithm is based on the fact that a combined image from the two cameras obtained at an OPD  
(i.e. the sum of the images from one interferogram) is a classical astronomical image (no interference fringes are visible), 
and that the combined images recorded right before and after this OPD are all very similar. 
This way, all sets of three successive combined images are compared to detect any abnormal variation of the flux at a given pixel. 
Once detected, cosmic rays are corrected by estimating the flux from the gaussian average of the neighboring pixels.
 
\item Sky Transparency Variations -  SITELLE's tilted design enable the measurements of the source 
intensity variations through time during the datacube acquisition. If the flux of the source itself does not change (which is the case 
for the vast majority of the astrophysical sources), variations of the combined flux recorded on the two cameras must come from variations 
of the atmospheric transmission (airmass or clouds) that can then be corrected for. 
A typical example of the atmospheric transmission function measured during the acquisition of a datacube on the center of M31 
observed through the SN3 filter is reproduced in Figure~\ref{fig:atm_trans}. 
This transmission function can also be compared with the one provided by the CFHT SkyProbe, co-aligned with the telescope. 

\begin{figure}
 \centering
 \includegraphics[width=\linewidth]{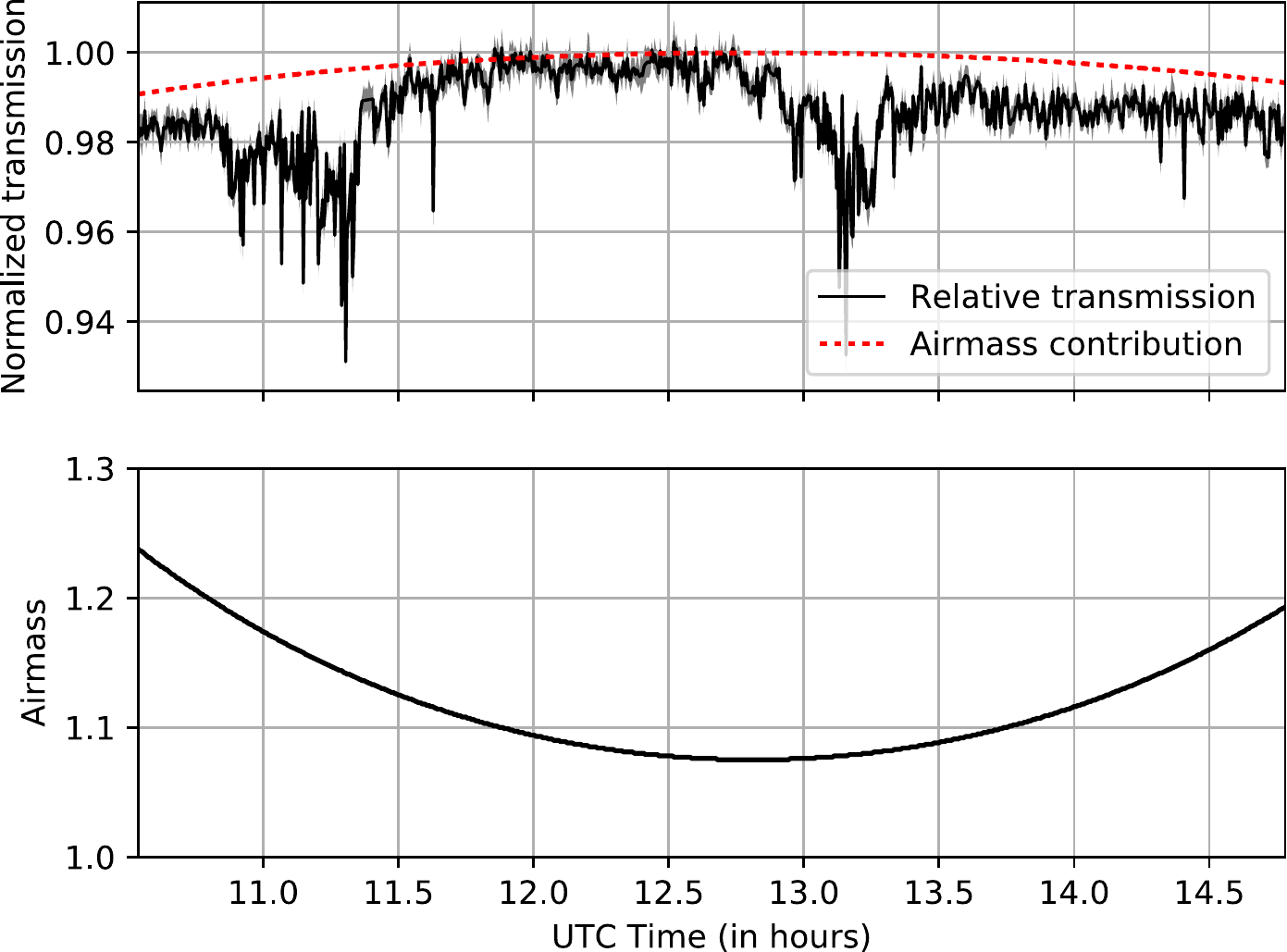}      
 \caption{
   \textit{Top:} Atmospheric transmission function during the acquisition of the M31 datacube in
   the SN3 filter observed on August 25, 2016 (courtesy of Anne-Laure
   Melchior, see \citealt{2018MNRAS.473.4130M}). The computed transmission is
   normalized to its 99th percentile. The grey surface represents the
   uncertainty. The airmass contribution to the atmospheric
   transmission is plotted in dotted red. It as been computed with a
   mean value of the extinction over the Mauna Kea at the H$\alpha$
   wavelength of 6.2\,$\times$\,10$^{-2}$\,mag/AM quoted from
   \citet{2013A&A...549A...8B}. \textit{Bottom:} Estimated airmass of the
   target.}
  \label{fig:atm_trans}
\end{figure}

\item  Fourier Transform and phase correction - The next step is to obtain the Fourier transform, based on a discrete Fourier transform (DFT) of each interferogram, of the combined interferometric datacube and to apply the phase correction. Phase errors and their correction,  resulting from a combination of factors including imperfect optics and interferograms sampling errors, will be discussed in details in Martin et al. 2018 (in prep).
The final product is a datacube in wavenumber (i.e. cm$^{-1}$).
 
\item  Wavelength and Flux Calibration - Wavelength calibration with SITELLE is secured using a high-resolution He-Ne (543\,nm) 
 laser datacube obtained at least once during the observing run. From the results obtained on several science targets, we have measured 
 small zero-point offsets from one datacube to the next, caused in part by the difference in the gravity vector at the position of the science target 
 and the laser calibration datacube. However, as mentioned in Section~\ref{sec:data_analysis}, 
using the night-sky OH lines present in the SN3 filter allows an excellent absolute calibration.  
 
Flux calibration comes from two different sources. The relative - wavelength dependant - calibration is computed from the acquisition 
of a spectrophotometric standard star datacube for each filter and obtained each semester. 
Once the wavelength dependency has been corrected, the absolute calibration is tied to a set of standard stars images taken 
in the vicinity of the science datacube. The atmospheric extinction is taken into account both within the science datacube 
and the standard images with an airmass dependent extinction based on the Mauna Kea sky transparency model of \citet{2013A&A...549A...8B}. 
The modulation efficiency measured from the standard star datacube is considered to be the same as for the science target datacube. 
\end{enumerate}

\begin{table*}
	\centering
	\caption{Datacubes discussed in this paper}
	\label{tab:targets_table}
	\begin{tabular}{lccccl}
		\hline
		Object &  Filter & R& Exposure/step & Num. Steps& Date\\
		\hline
		M57 & SN3 & 2650 & 15s & 593 & August 8, 2015 \\
		M1 & SN3 & 9500 & 5.3s & 1786 & Nov 22, 2016 \\
		Orion & SN1 & 3500 & 10s & 1500 & Jan 30, 2016 \\
		M1-71 & SN3 & 1500 & 8s & 269 & August 10, 2015 \\
		M51 & SN3 & 1490 & 40s & 251 & May 9, 2016 \\
		M33 & SN3 & 3000 & 18s & 505 & October 13, 2017 \\
		M33 & SN2 & 1020 & 38s & 219 & September 28, 2017 \\
		M33 & SN1 & 1020 & 49s & 172 & September 28, 2017 \\
		Arp 94 & SN3 & 1250 & 11s & 223 & January 12, 2016 \\
		Arp 94 & SN2 & 600 & 47s & 138 & January 12, 2016 \\
		Arp 94 & C1 & 600 & 33s & 389 & January 10, 2016 \\
		Abell 168 & SN3 & 1000 & 35s & 186 & August 10, 2015 \\
		COSMOS & SN2 & 600 & 102s & 135 & May 9, 2016 \\
		Abell 2390 & C4 & 1300 & 92s & 124 & July 6, 2017 \\

		\hline
	\end{tabular}
\end{table*}

\subsubsection{Data Analysis}
\label{sec:data_analysis}

Given the size of the spectral datacubes (up to 34\,Go), the unusual instrumental line shape (ILS), and the very
large number of spectra to analyze, a specific data analysis library (ORCS)
has been developed to help the user to handle SITELLE's datacube. 
Different tools are proposed which have already been used for
the analysis of different astrophysical objects \citep{2017MNRAS.465..739S,
  2016MNRAS.463.4223M, 2018MNRAS.473.4130M, 2018MNRAS.479L..28G,2018MNRAS.tmp..465R}; they include: 
  
\begin{enumerate}
\item Model fitting - Emission line parameters can be extractred
  through model fitting, and flux and velocity maps of the
  lines can be generated. The fitting engine has been parallelized to fit the large
  number of spectra (4 million) for each datacube. Different emission
  line models are implemented. All maps and spectral fits shown in  Section 5 have been
  generated with ORCS.
  
\item Calibration - The wavenumber calibration can be refined by
  fitting the OH sky lines with ORCS, which are quite strong in the SN3 filter (e.g. see Figure~9). An absolute precision of about 
  1\,-\,3\,km\,s$^{-1}$ may be reached.

\end{enumerate}

\section{Instrument Capabilities and a Highlights of Some Results}

Science results published so far include the kinematics and ionisation structure of the planetary nebula M57 \citep{2016MNRAS.463.4223M}, the kinematics of the nova shell around AT Cnc \citep{2017MNRAS.465..739S}, detection of $\sim$ 800 emission-line point sources in the bulge of M31 \citep{2018MNRAS.473.4130M}, an analysis of the physical properties of more than 4000 HII regions in the spiral galaxy NGC 628 \citep{2018MNRAS.tmp..465R}, as well as the complete mapping of the nebula associated with NGC 1275 in the Perseus cluster \citep{2018MNRAS.479L..28G}. A number of 
datacubes were also obtained during SITELLE commissionning and later in order to assess its capabilities. We present here some of these results to highlight SITELLE's capabilities, leaving the full analysis to subsequent papers. Table~\ref{tab:targets_table} contains the observing information for these data.

\subsection{Instrument Line Shape}

The bright planetary nebula M57 was an ideal target for SITELLE's first light, in particular to assess the quality of the Instrument Line Shape (ILS): 
M57 shows a very bright core, allowing us to accurately measure the ILS, but also a faint, highly structured halo with strongly varying line ratios. A detailed analysis of the M57 data, with an emphasis on nebular kinematics, is presented in \citet{2016MNRAS.463.4223M}.

In the ideal case of a perfectly monochromatic source observed for an infinite amount of time, an FTS's ILS would be a delta function (the Fourier transform of an infinite sinus).
But as a result of the finite size of the interferogram, an FTS ILS is not a simple gaussian (a widened delta function) but rather the Fourier transform of a boxcar function (which defines the beginning and the end of the data collection): a cardinal sine, or sinc function. 
Any departure from an ideal sinc function caused by an incorrect sampling of the interferogram (e. g. due to errors in the metrology) would be readily observed. 
On the other hand, an intrinsic broadening of the natural line shape from the source, caused for instance by internal morions of the gas, will result in a convolution of the sinc function with a gaussian. This function, named sincgauss, has been discussed in \citet{2016MNRAS.463.4223M} and implemented in ORCS. 
\\
\\
 
\begin{figure}
	\includegraphics[width=\columnwidth]{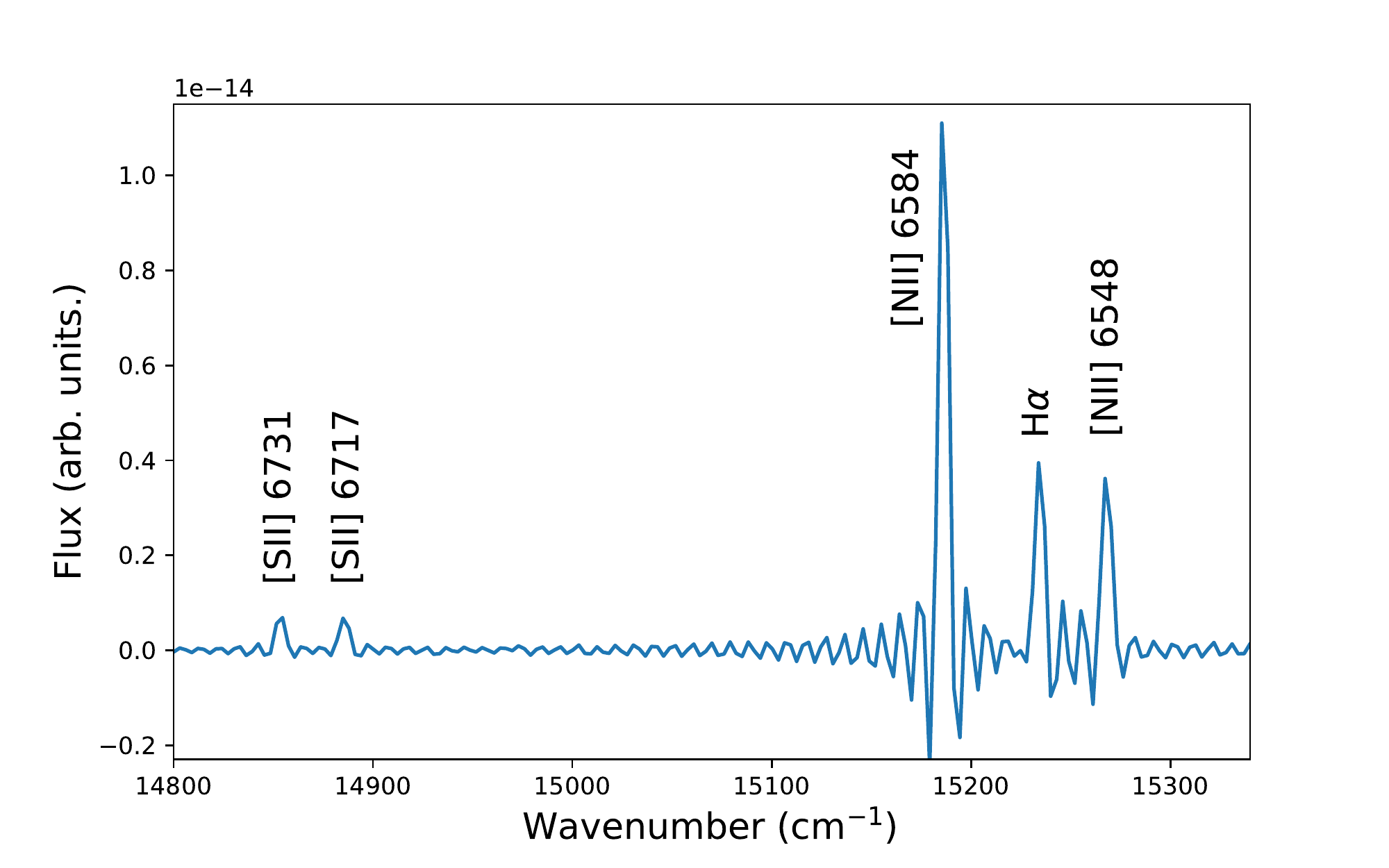}
    \caption{Spectrum from the bright part of M57, showing the well-defined sinc profile characteristic of the FTS.}
    \label{fig:M57spec}
\end{figure}

\begin{figure}
	\includegraphics[width=\columnwidth]{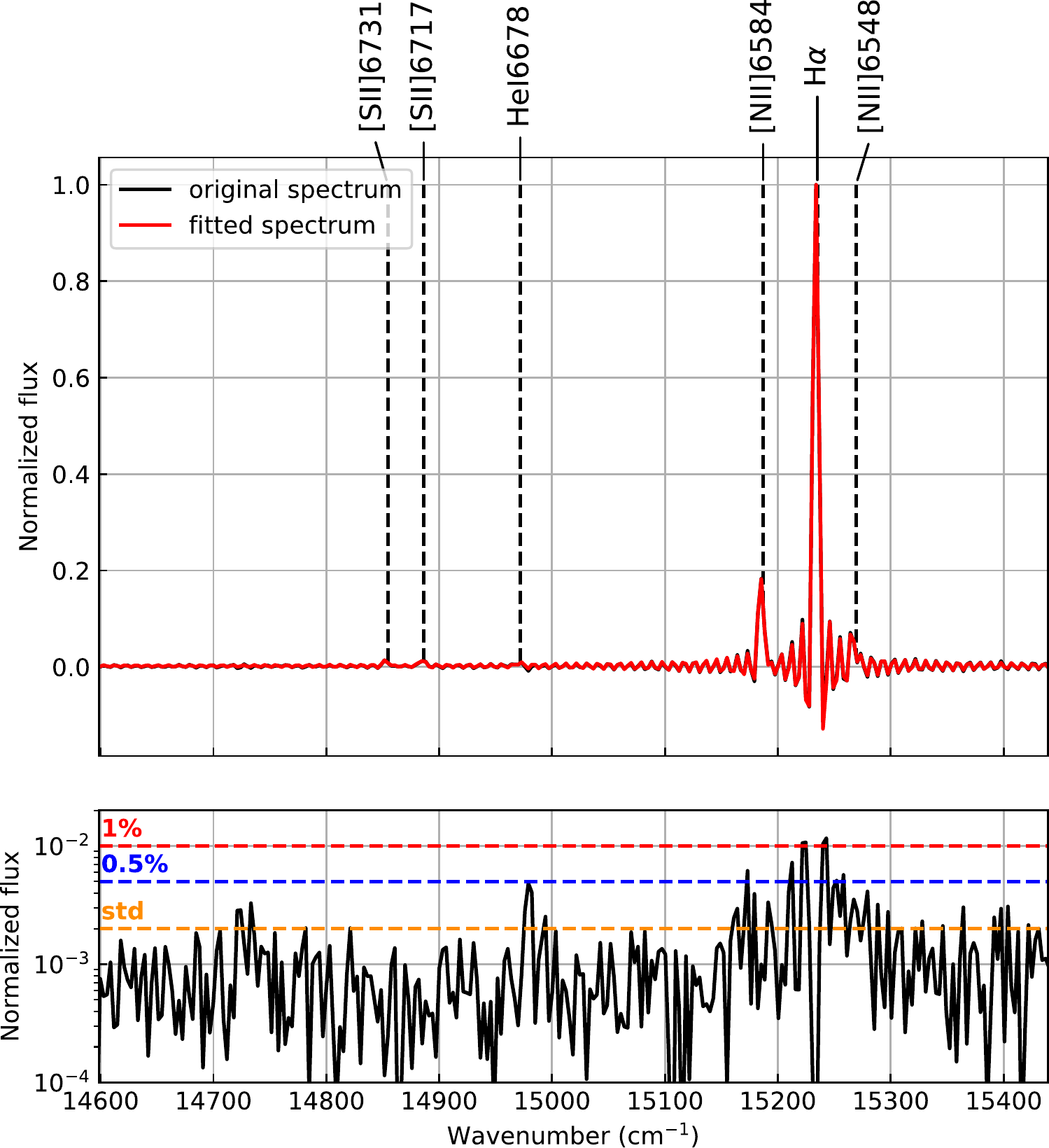}
    \caption{ORCS fit on an unbinned spectrum of M57 (upper panel) and residuals (lower panel). 
    }
    \label{fig:Residuals}
\end{figure}

A spectrum from the bright part of the nebula (Figure~\ref{fig:M57spec}) in the SN3 filter, including 5 bright lines (and the much fainter He{\sevensize I}\,$\lambda$6678) illustrates SITELLE's typical instrument line function while Figure~\ref{fig:Residuals} presents a fit with very small residuals, attesting to SITELLE's high quality ILS.

Obviously, the ILS sidelobes of the brightest lines, which extend all the way to the edge of the filter bandpass, significantly affect the visual intensity and shape of the fainter lines. It is thus essential to fit all lines simultaneously, with the proper sinc function, in order to extract the correct flux, wavelength, and an eventual enlargement of all lines. This is the purpose of ORCS, SITELLE's data analysis software described  in \citet{2016MNRAS.463.4223M}.

\subsection{Spectral Resolution}
The spectral resolution of an iFTS datacube is defined by the maximum OPD, or twice the maximum physical displacement of the moving mirror from its original position (ZPD).
The maximum theoretical spectral resolution provided by SITELLE's interferometer is therefore set by the maximal mechanical displacement of the moving mirror (5\,mm) and varies from R\,=\,18\,000 at 900\,nm up to 47\,000 at 350\,nm.
Several practical constraints however significantly lower these values.  First, the point spread function (set by the optics and the seeing) lowers the maximum resolution reachable at very high values of R because the separation between interference fringes becomes smaller than the optical resolution; it does not, however, decrease the resolution for typical datacubes (R\,$\leq$\,5000) because the fringes are wider than the seeing. Also, the separation between the fringes is not uniform across the FOV (because of the off-axis configuration) and the contrast between fringes diminishes as the OPD increases. But more importantly and practically, the low surface brightness of most astronomical sources naturally limits the number of steps, and therefore spectral resolution, in order to collect enough photons at each step while maintaining a reasonable total observing time and minimizing the overheads. We recall that the step size (and hence the number of steps required to reach the maximum OPD; see Table~\ref{tab:filters_table}) is set by the Nyquist criterion. For example, if we consider a datacube with the SN3 filter limited to four hours, the overhead (CCD readout time of 3.8\,s per step) is only 4.7\% of the on-target integration time for R\,=\,1000 while it goes up to 29\% at R\,=\,5000. Another point is worth mentioning: while in principle it would be sufficient, in order to reach a given spectral resolution, to start the interferograms at the ZPD and move the mirror until we reach the required OPD, we prefer to start the datacube before the ZPD, at a distance corresponding to 25\% of the required maximum OPD, in order to better define the phase correction to be applied to the data (much more details on phase correction are presented in Martin \& Drissen 2018, in prep.).

In order to test SITELLE's spectral capabilities, we have obtained a datacube of the Crab supernova remnant using the SN3 filter, aiming 
for R\,=\,10\,000. Figure~\ref{fig:M1sky} shows the spectrum of a $30'' \times 30''$ region outside the nebula, dominated by night-sky OH lines. A fit using ORCS, to this spectrum as well as others in the field, shows that an average spectral resolution of R\,=\,9500 was reached. This value is about only 5\% smaller than that expected from the maximum OPD reached by the Michelson interferometer. As expected, all other datacubes obtained at lower values of R so far with SITELLE, at all wavebands, reached the predicted spectral resolution.

\begin{figure*}
	\includegraphics[width=5.5cm]{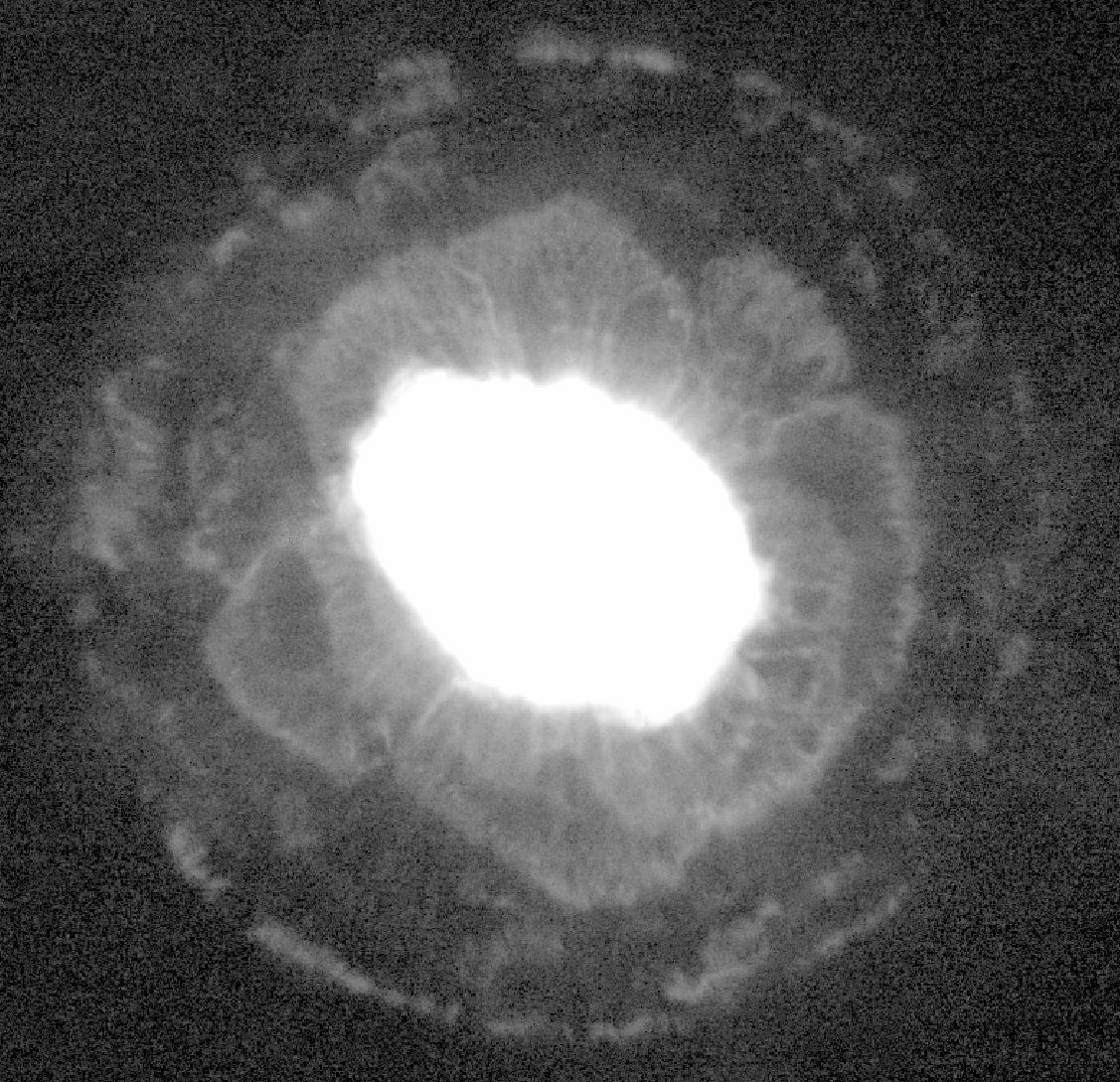}
		\includegraphics[width=5.5cm]{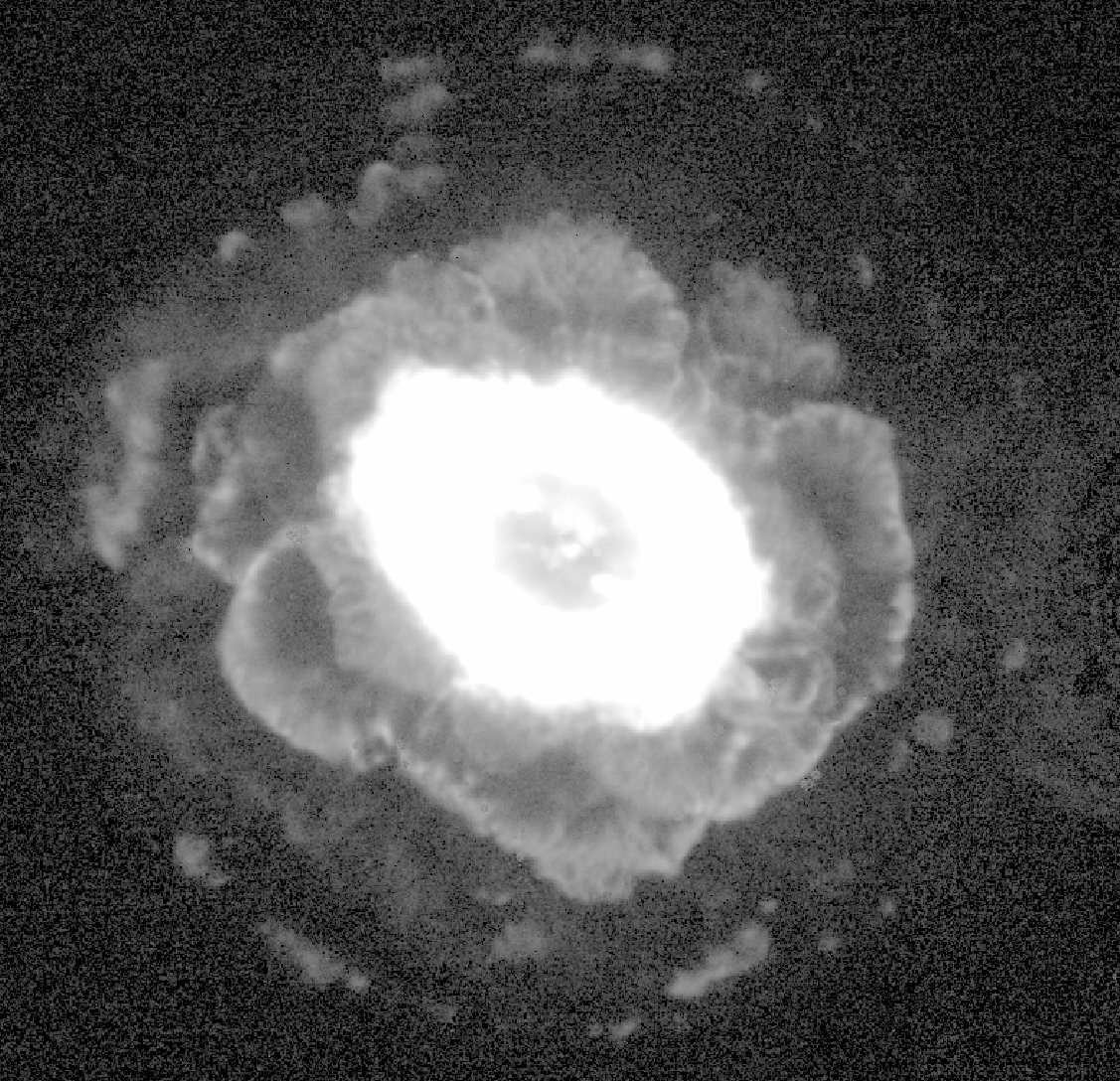}
			\includegraphics[width=5.32cm]{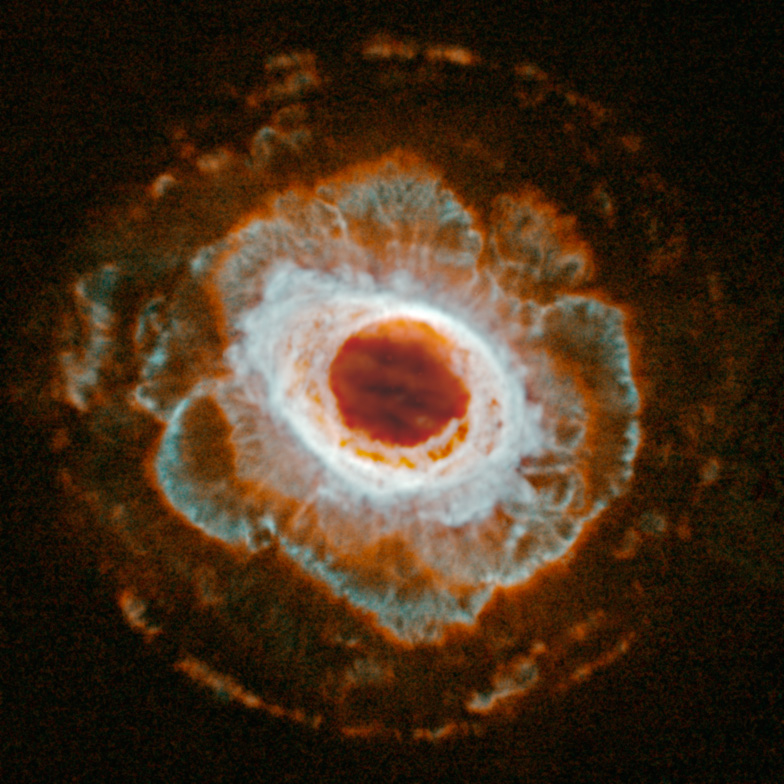}
    \caption{Pure H$\alpha$ (left), [N\II]\,$\lambda$6584 (middle), and composite images of M57 extracted from the SITELLE's SN3 datacube. The bright central region has been saturated on the first two images to enhance the faint, outer halo, but is safely below saturation in the datacube itself as shown in the composite image (right). FOV is $4' \times 4'$ (about 15\% of the entire SITELLE's FOV), with North at the top and East to the left.}
    \label{fig:M57Ha}
\end{figure*}

\begin{figure}
	\includegraphics[width=\columnwidth]{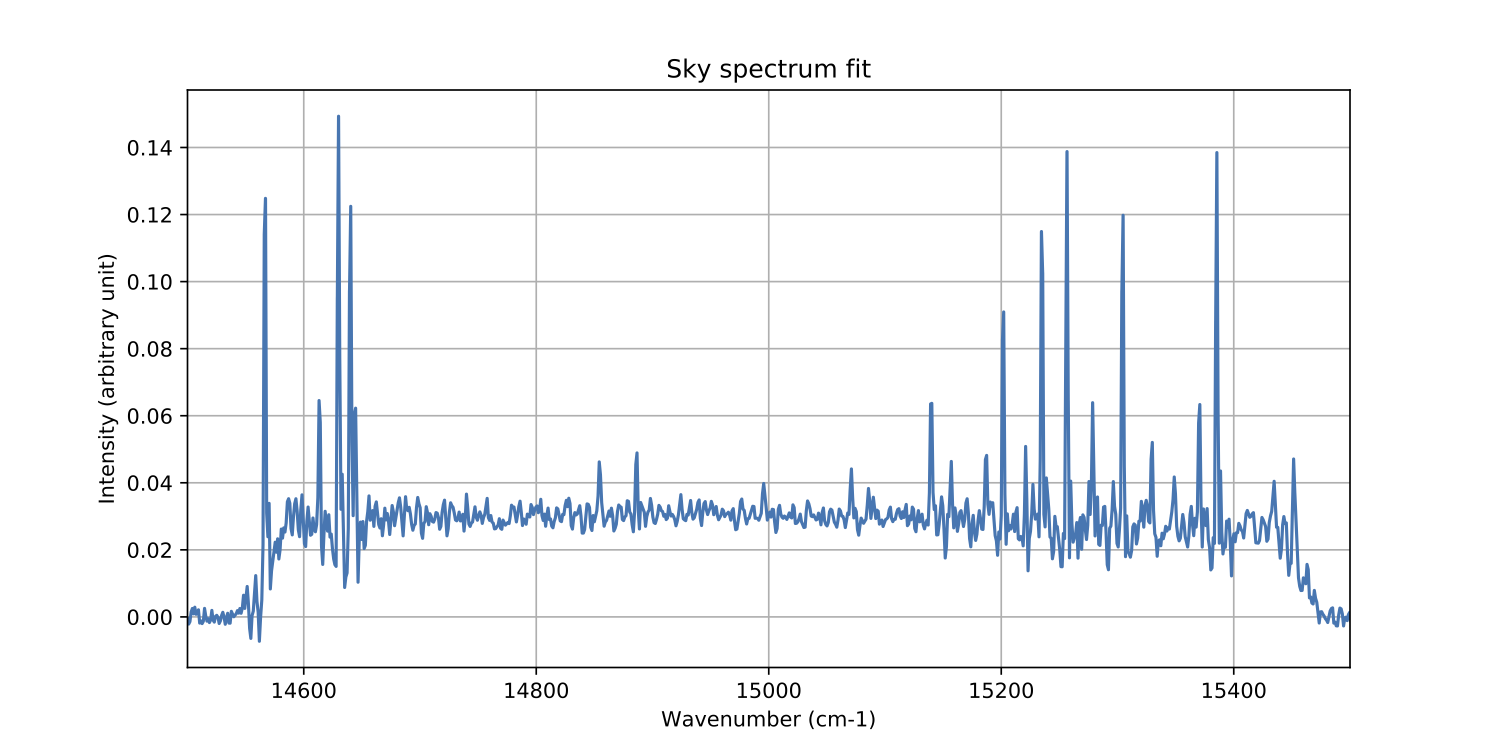}
	\includegraphics[width=\columnwidth]{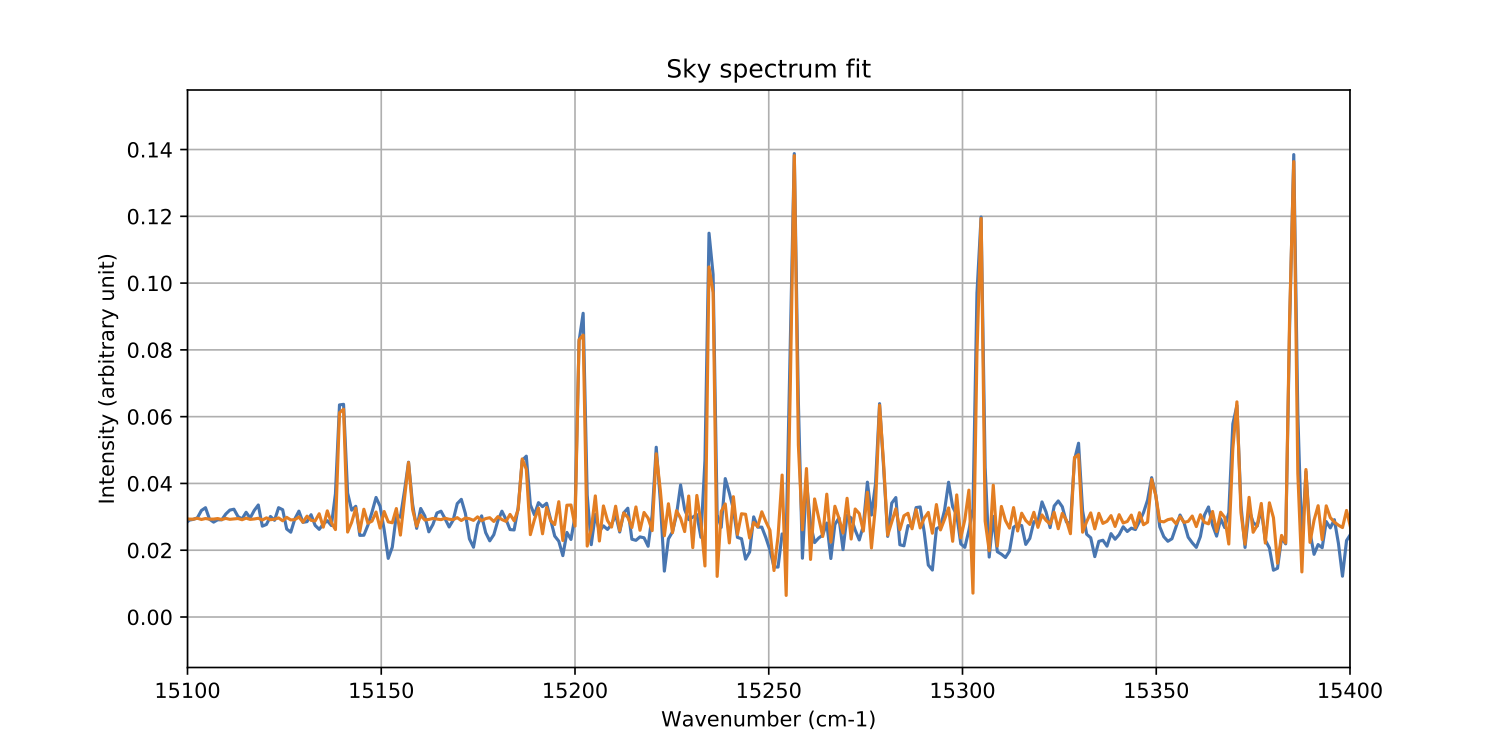}
    \caption{\textit{Upper panel}: Integrated spectrum of a $30'' \times 30''$ region of the night sky besides the Crab nebula in the SN3 datacube; the edges of the filter are clearly visible. \textit{Lower panel}: Enlargement of the blue section of the spectrum, with fits to the night sky OH lines using ORCS. The average spectral resolution is R\,=\,9500.}
    \label{fig:M1sky}
\end{figure}

Figure~\ref{fig:M57Ha} shows monochromatic and composite images in H$\alpha$ and [N\II]\,$\lambda$6584 of the entire nebula extracted using the ORCS library. 

\begin{figure}
  \includegraphics[width=\columnwidth]{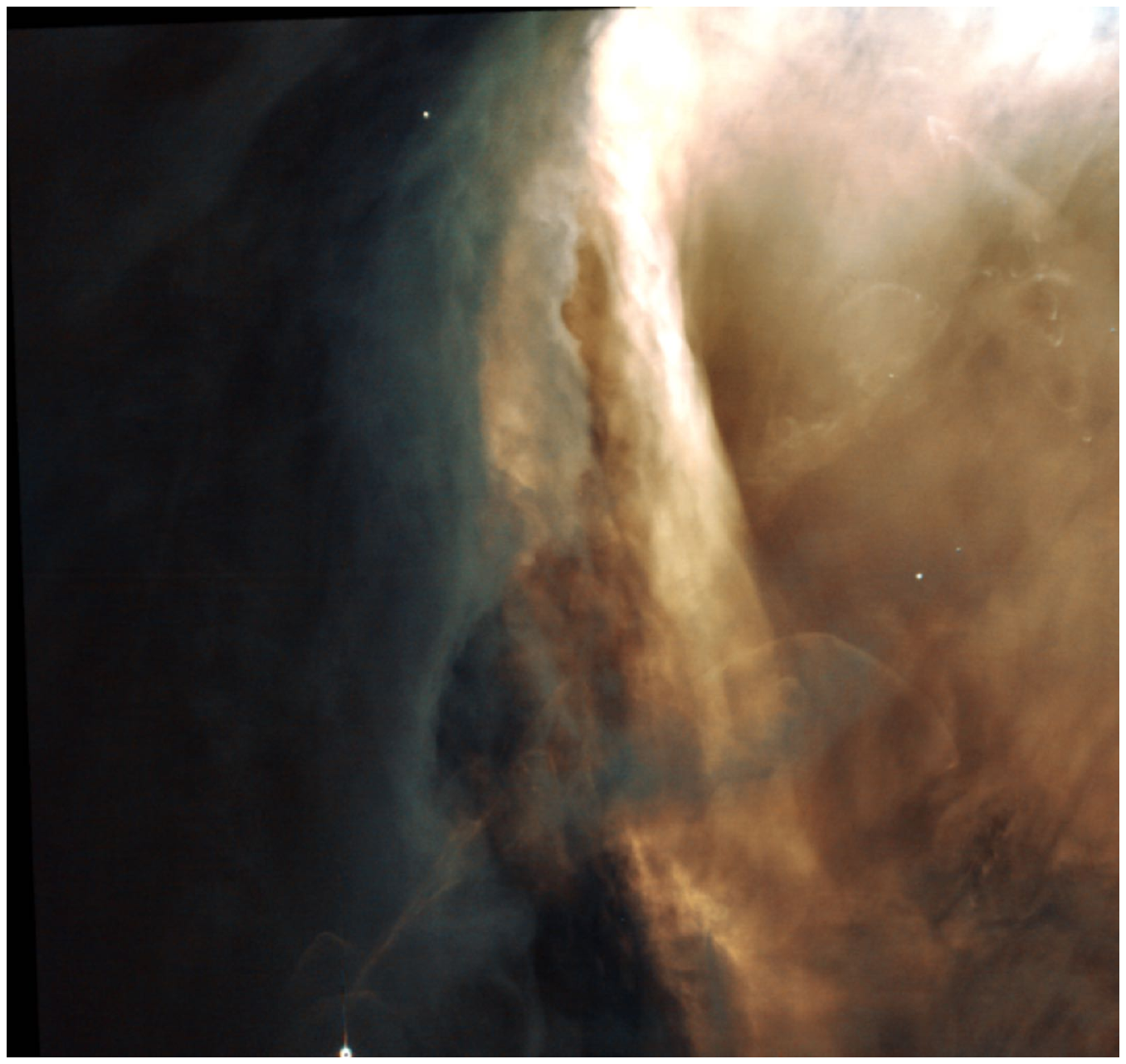}
    \includegraphics[width=\columnwidth]{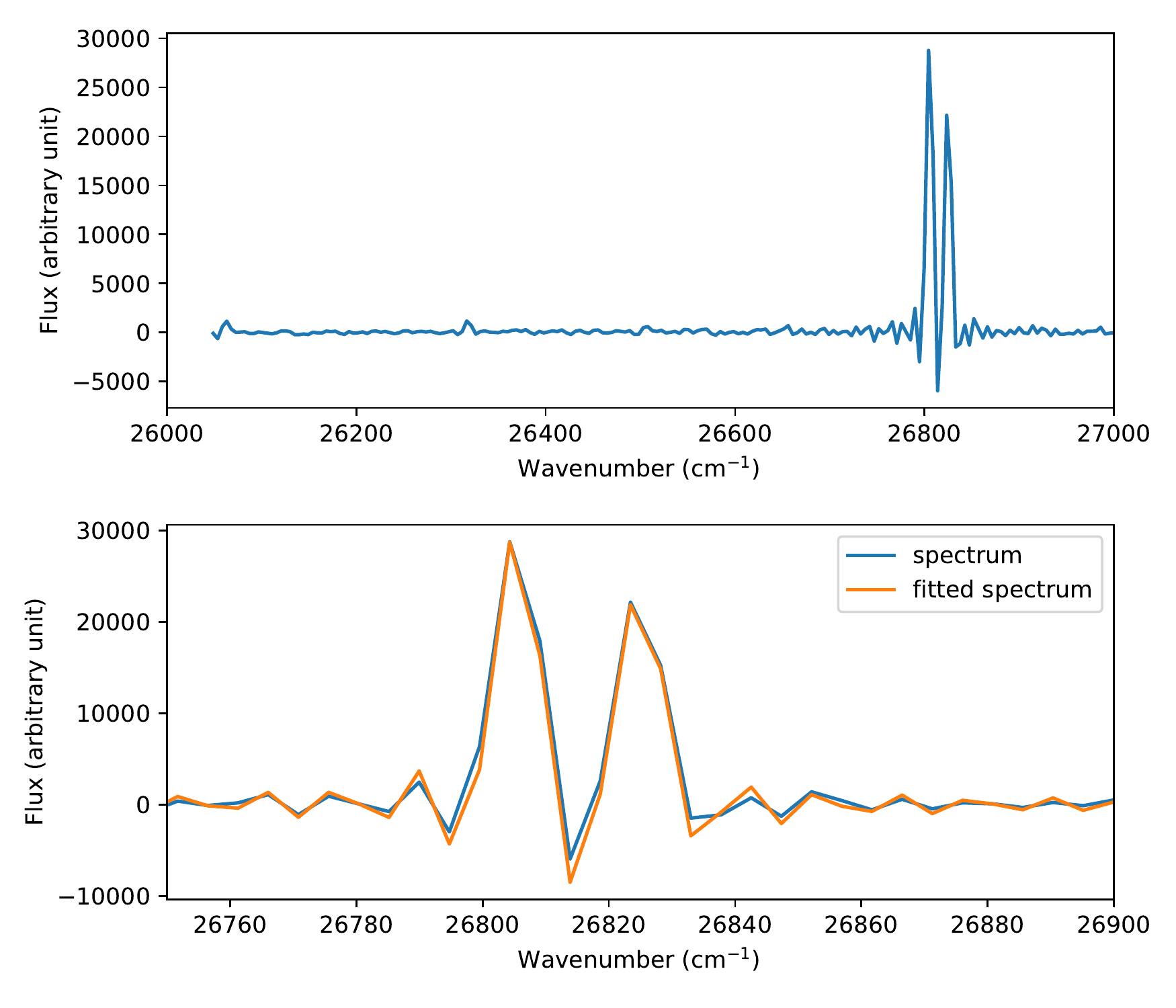}
  \caption{\textit{Upper panel}:  Luminosity-weighted electron density image of a region in the Orion nebula ($11' \times 11'$ centered on 05h35m44.2s, $-$5$^\circ$33$'$00$''$) based on the ratio of the [O\II]\,$\lambda\lambda$3726,3729 doublet; blue hue corresponds to regions of higher dentity. North is at the top, East to the left. \textit{Middle panel}: Spectrum of a 2\,arcsec$^2$ region showing the entire SN1 wavelength range. 
\textit{Lower panel}: Enlargement of the same spectrum centered on the [O\II] doublet with a fit using ORCS.
}
    \label{fig:OrionOII}
\end{figure}

One of the main technological challenges behind SITELLE was the need to reach the [O\II]\,$\lambda$3727 doublet. Not only did it require the optics' transmittance and the detector's quantum efficiency to be high, which is relatively easy to reach nowadays, but it imposed severe constraints on the interferometer's properties. We have obtained an SN1 datacube of the Orion nebula with the aim of pushing the resolution high enough to separate the two components of the [O\II] doublet; although this requires a much higher spectral resolution than the separation of the [S\II]\,$\lambda\lambda$6717,6731 doublet with the SN3 filter, [O\II] is intrinsically ten times brighter \citep{2007A&A...465..207S}.  Beyond testing SITELLE with an extended, nebulous, filling the FOV object in the near-UV spectral domain, one of the scientific goals behind the observation of Orion was to obtain spectra of 
compare the electronic densities derived from the [O\II] and [S\II] line ratios. Since these ions have different ionization potential 
(separated by 3.2\,eV), the comparison of the two density maps would determine if the difference is enough to offer the 
characterization of two different ionization volumes.
Figure~\ref{fig:OrionOII}  shows the intensity map of the doublet, color-coded according to the [O\II]\,$\lambda\lambda$3726,3729 doublet ratio,
 as a proxy for the electron density; a quantitative analysis of these data will be presented elsewhere (Joncas et al., in preparation).
A typical spectrum extracted from the datacube (Figure~\ref{fig:OrionOII}; middle panel) shows that the doublet is indeed clearly separated, and a fit with ORCS (Figure~\ref{fig:OrionOII}; lower panel) confirms that the goal of R\,=\,3800 was reached. More recently, the highest spectral resolution datacube attempted with the SN1 filter was R\,$\simeq$\,9000, aiming at studying the kinematics and electron density distribution of the Eagle Nebula (M16; Flagey et al., in preparation). 

\subsection{Image Quality}
\begin{figure}
  \includegraphics[width=\columnwidth]{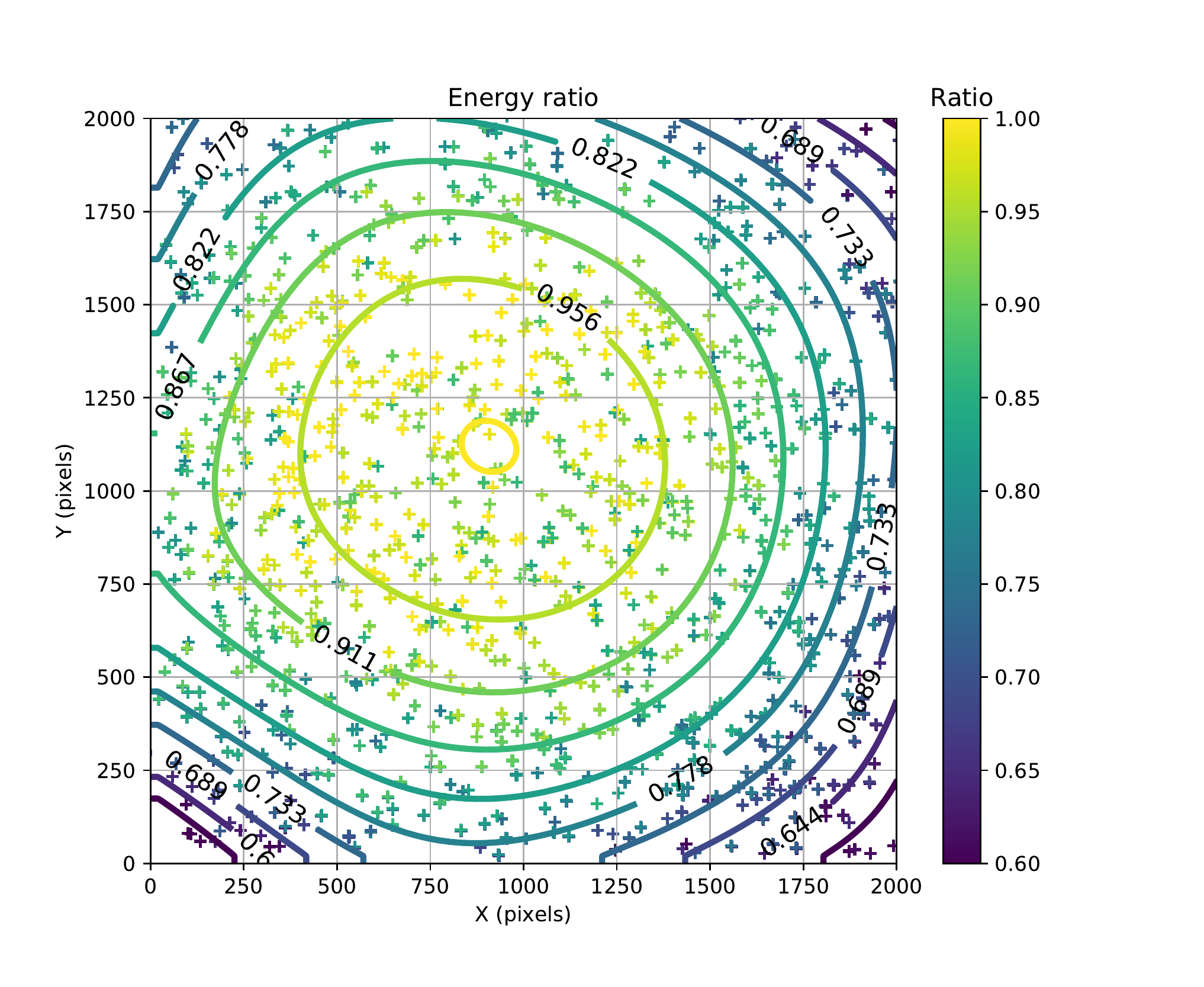}
  \caption{Relative ratio of encircled energy across the field of view (see text).}
    \label{fig:Encircled}
\end{figure}

Very early in the commissioning phase, it was noticed that the image quality at the edge of the FOV, 
in particular in the upper corners, was not nominal.
We have evaluated the amount of energy lost in the wings of the point spread function (PSF) of the stars in
the corners of the FOV by measuring, for a large number of
point-sources, the ratio of the energy contained in a circular
aperture of diameter 1\arcsec\ diameter over the energy contained in a circular
aperture of  3.8\arcsec\  diameter. About 10\,000 stars have been
selected in three images of the same galactic field around the nebula
M1-67 obtained with three different filters: SN3, SN2, and C2. As the seeing
and the PSF are different from one image to the other, the calculated
ratios have then been normalized to the highest measured ratio in each
image. The resulting values shown in Figure~\ref{fig:Encircled} thus reflect the
relative loss of energy in the wings of the PSF with respect to the
central region of the FOV where the PSF is the pointiest. The contours
come from a 2D spline model fitted to these values. Work is in progress in order to understand the source
of the problem and propose a solution.

\subsection{Complementary Interferograms and Deep Images}
\label{sec:deep}
\begin{figure*}
  \includegraphics[width=\columnwidth]{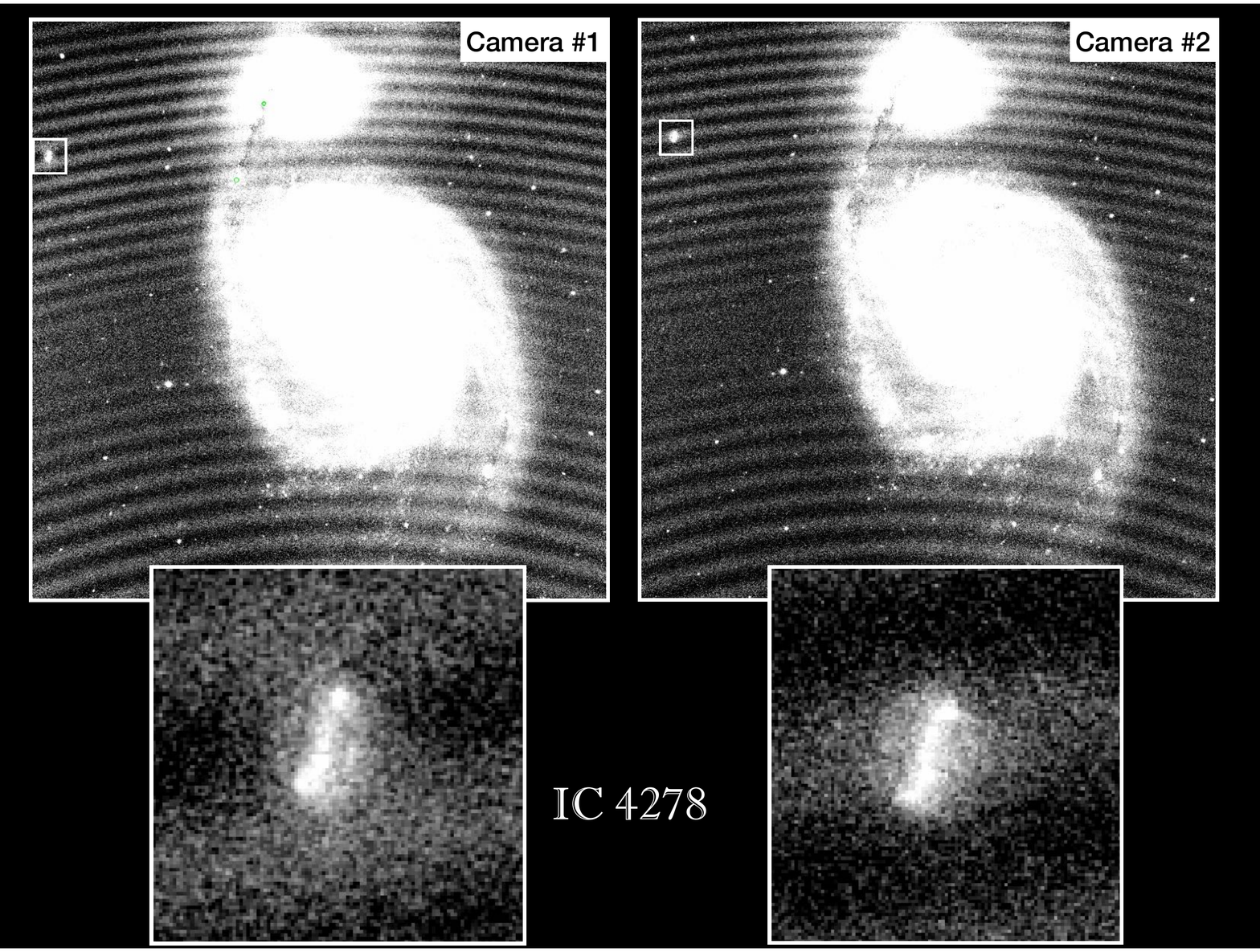}   
   \includegraphics[width=6.25cm]{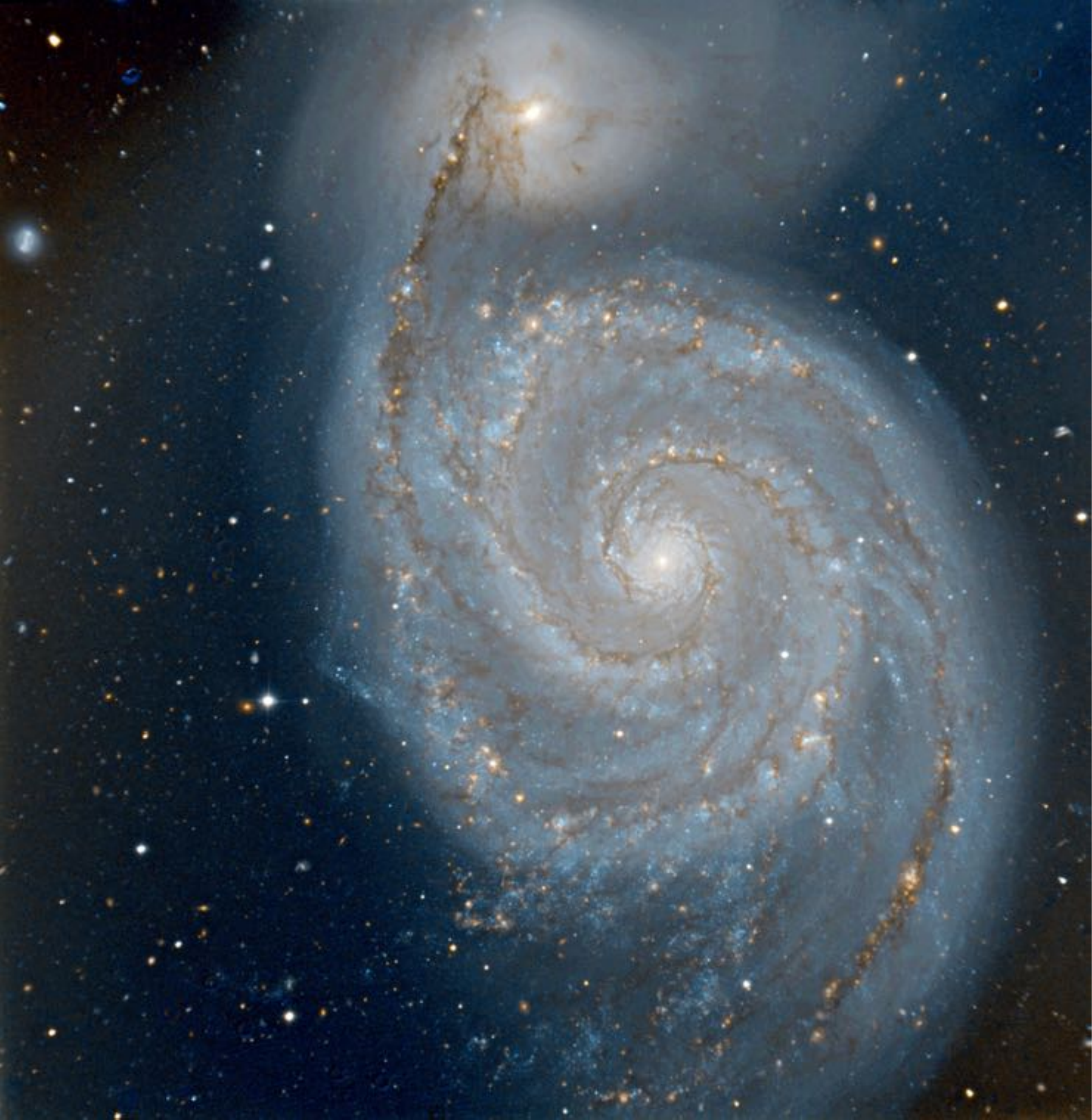}
  \caption{\textit{Left}: Images extracted from the raw interferogram cubes, showing the complementary fringe patterns in the two cameras. These data of M51 were obtained with the SN3 filter. The \textit{inserts}, centered on the background galaxy IC\,4278, clearly show the presence of complementary fringes on the two outputs. \textit{Right}: Color-coded combination of the deep SN2 and SN3 images of M51, each one resulting from the co-addition of all images, in both cameras, from the interferometric cube. FOV is $10.5' \times 10.5'$ with North at the top and East to the left.}
    \label{fig:M51a}
\end{figure*}

\begin{figure}
     \includegraphics[width=8.5cm]{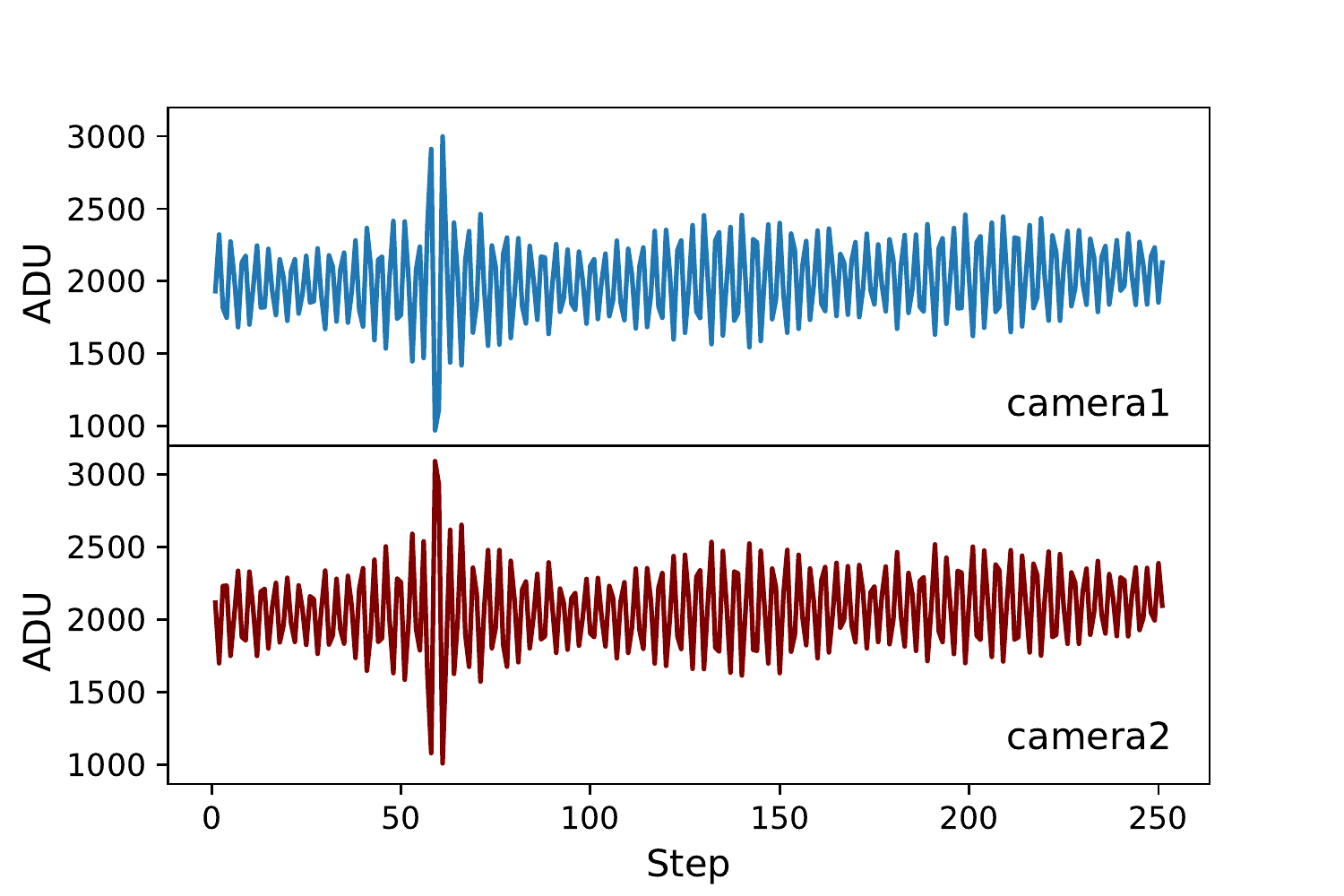}
          \includegraphics[width=8.5cm]{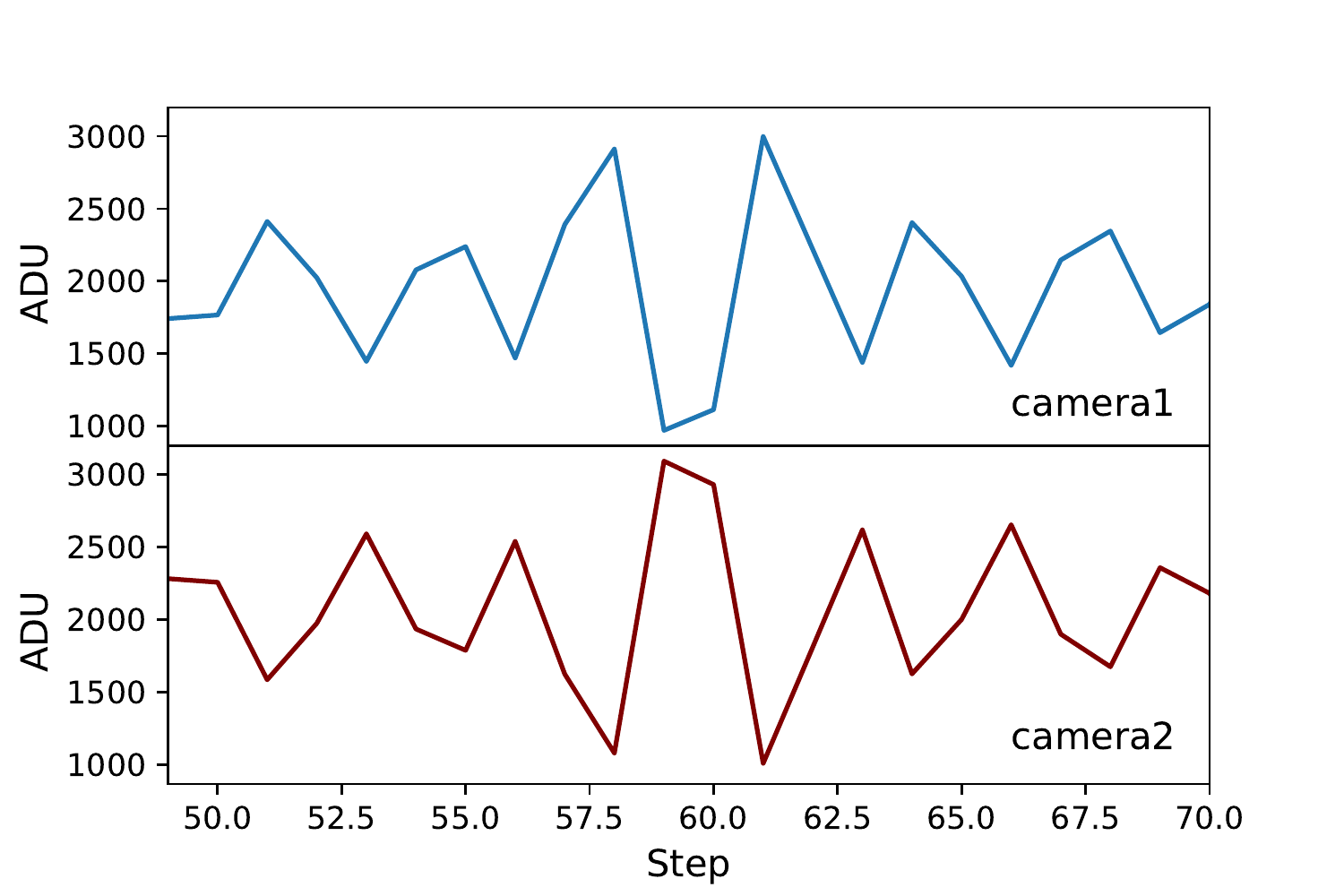}
          \includegraphics[width=8.5cm]{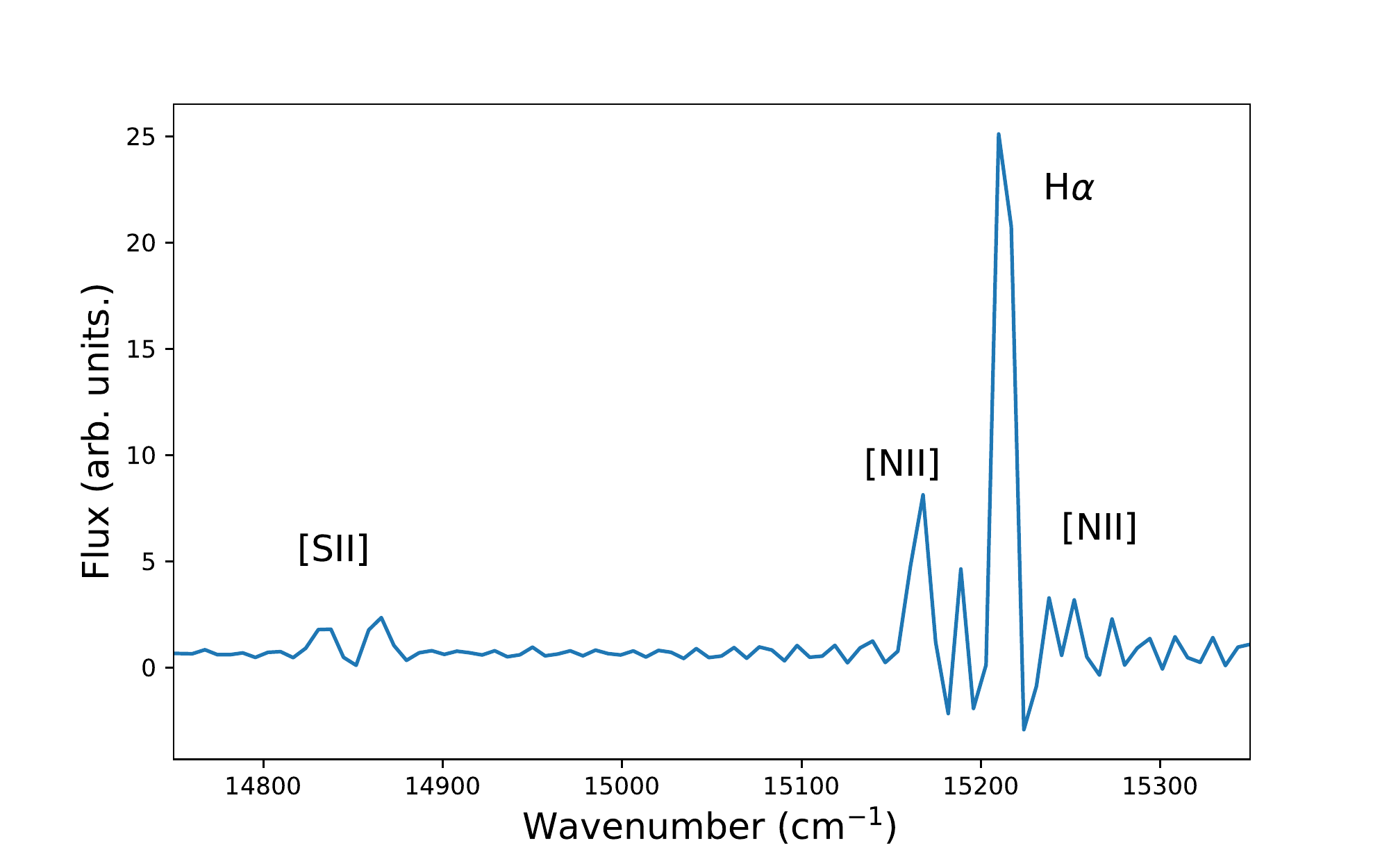}
  \caption{\textit{Upper panel}: Interferograms of a bright H\II\ region in M51 in both cameras. The main beating pattern seen in the interferograms is caused by the interaction between the [N\II] and H$\alpha$ lines. A closer look at this figure shows that when a strong signal is detected in one camera, a low signal is detected in the other. This is particularly obvious around the ZPD (around step 60; \textit{middle panel}), corresponding to equal optical path differences in the two arms of the interferometer.  \textit{Lower panel}: Spectrum of the same region after a complete treatment of the above interferograms with ORBS.}
    \label{fig:M51b}
\end{figure}

The raw images obtained by SITELLE's cameras at each mirror step are complementary interference patterns. The sum of these two images is thus equivalent to an image obtained with a single camera without the Michelson interferometer.
Therefore, as a by-product of the datacube, a deep image is produced by co-adding all the interferograms from both cameras. Because the readout noise is low compared to the photon noise from the combination of the night sky and the object, this deep image is very similar to a single, long exposure of the target with an exposure time equal to the sum of the individual exposures of the datacube. Figure~\ref{fig:M51a} illustrates this process with images extracted from a raw interferometric cube of M51 obtained with the SN3 filter. Because the bandpass covered by this filter includes night-sky OH lines, constructive and destructive fringes are seen across the entire FOV. A close look at these images clearly shows the complementary nature of these fringes from one camera to the other. An even closer look at a bright individual H\II\ region in the galaxy would reveal a different fringe pattern due to the significantly different spectral content of the sky and the H\II\ region; in fact, both fringe patterns would then be sumperimposed, the one produced by the H\II\ region clearly dominating. The deep image resulting from the combination of the interferograms of the SN3 and SN2 cubes is also shown in Figure~\ref{fig:M51a} (right panel). Figure~\ref{fig:M51b} shows interferograms of a bright H\II\ region in M51 (integrated over a radius of 2$''$) as recorded by both cameras, as well as the spectrum resulting from the data reduction. 

\subsection{Sensitivity}
\begin{figure*}
  \includegraphics[width=14cm]{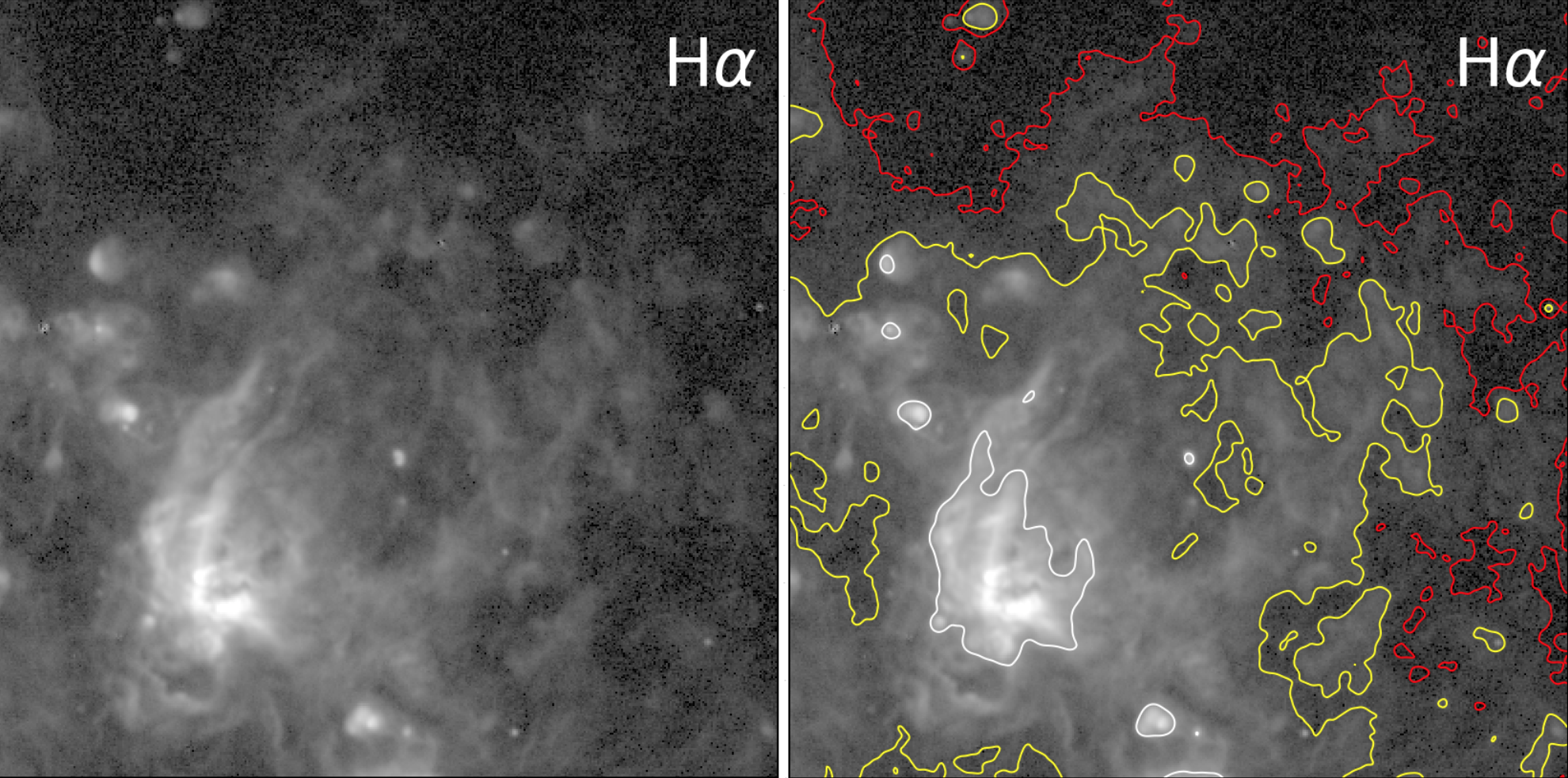}
    \includegraphics[width=14cm]{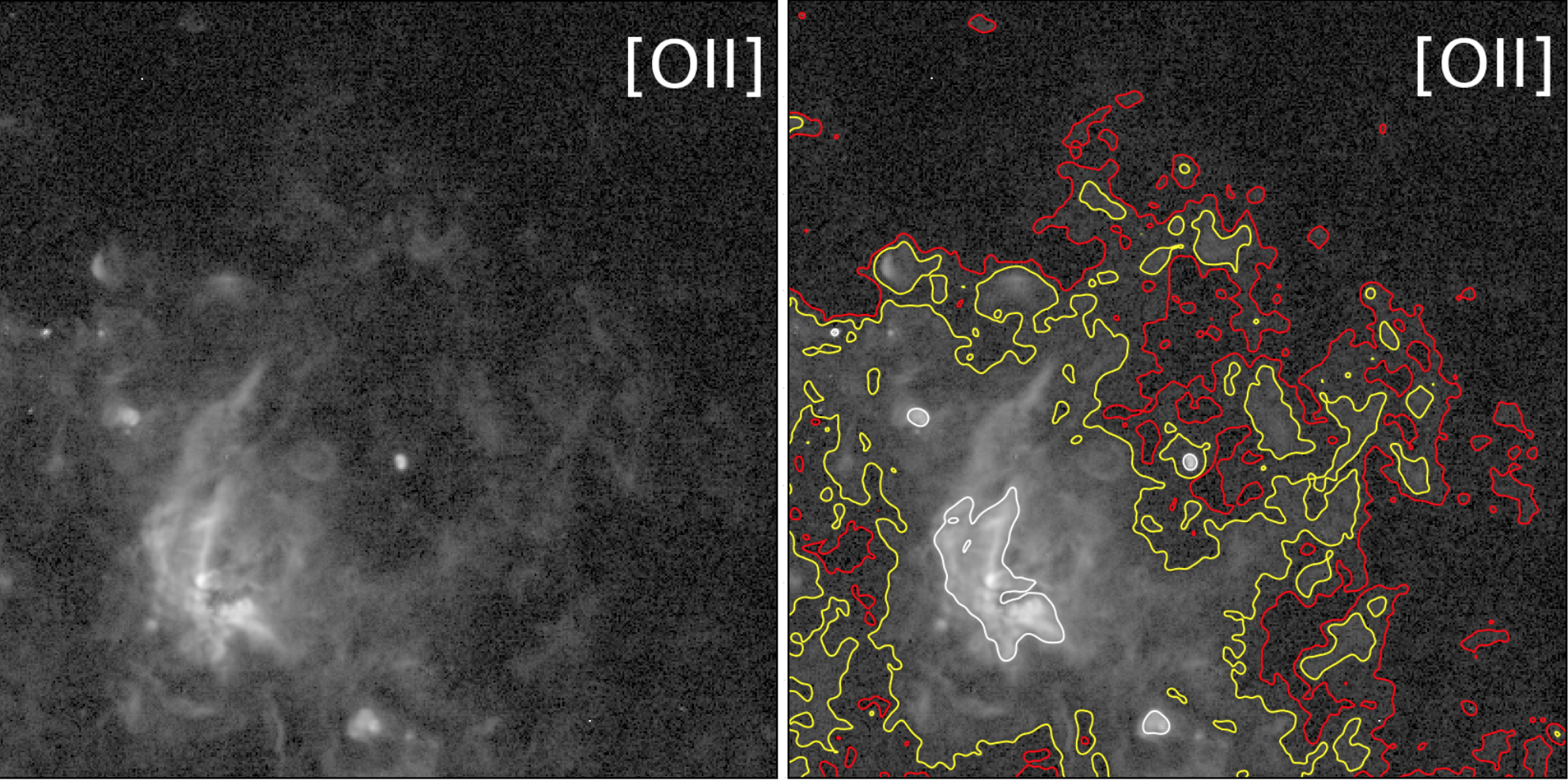}
  \caption{\textit{Upper panel}: H$\alpha$ map of the region around NGC\,595 in M33 extracted with ORCS from the SN3 datacube, using a $2 \times 2$ binning. Both images are identical, but contours (4, 10, and 100\,$\times$\,10$^{-17}$\,erg~cm$^{-2}$\,s$^{-1}$\,arcsec$^{-2}$ in red, yellow, and white, respectively) have been added on the right. \textit{Lower panel}:  Same as upper panel, but for the [O\II]\,$\lambda$3727 line, and with a lower contour of  
  7\,$\times$\,10$^{-17}$\,erg~cm$^{-2}$\,s$^{-1}$\,arcsec$^{-2}$  in red. FOV is 4.25$' \times 4.25'$, with North at the top and East to the left.}
    \label{fig:N595a}
\end{figure*}

Although several datacubes were obtained during SITELLE's commissioning and Science verification observing runs, the most homogeneous and best photometrically calibrated dataset was obtained in the fall of 2017: four fields in the center of the Local Group galaxy M33 were obtained in the SN1, SN2, and SN3 filters, with very similar observing time. Figure~\ref{fig:N595a} shows H$\alpha$ and [OII] maps of a region around the active star-forming region NGC 595 obtained with ORCS using a 2$\times$2 pixel binning. Diffuse components are detected at the 
4\,$\times$\,10$^{-17}$\,erg~cm$^{-2}$\,s$^{-1}$\,arcsec$^{-2}$ level in SN3 and 7\,$\times$\,10$^{-17}$\,erg~cm$^{-2}$\,s$^{-1}$\,arcsec$^{-2}$ in SN1. Similar values are obtained with the SN2 filter.

\subsection{M1-71: Spectrophotometric Calibration and Faint Structures}

M1-71 (PNG\,055.5$-$00.5) is a compact, elongated planetary nebula that was observed for calibration purposes during  SITELLE commissioning.
Although very little is known about its internal structure and kinematics, nor its distance, its H$\alpha$ flux has been accurately measured using wide-aperture spectroscopy  \citep{2005A&A...436..967W} and, more recently, narrow-band imagery \citep{2013MNRAS.431....2F}; both are in excellent agreement with each other. We find an integrated H$\alpha$ flux 8\% lower than these previous measurements. This difference could in part be caused by the assumed value of the modulation efficiency (ME) during the observations; we did not, during commissioning, have an optimized strategy to accurately measure this value. 

\begin{figure*}
	\includegraphics[width=\columnwidth]{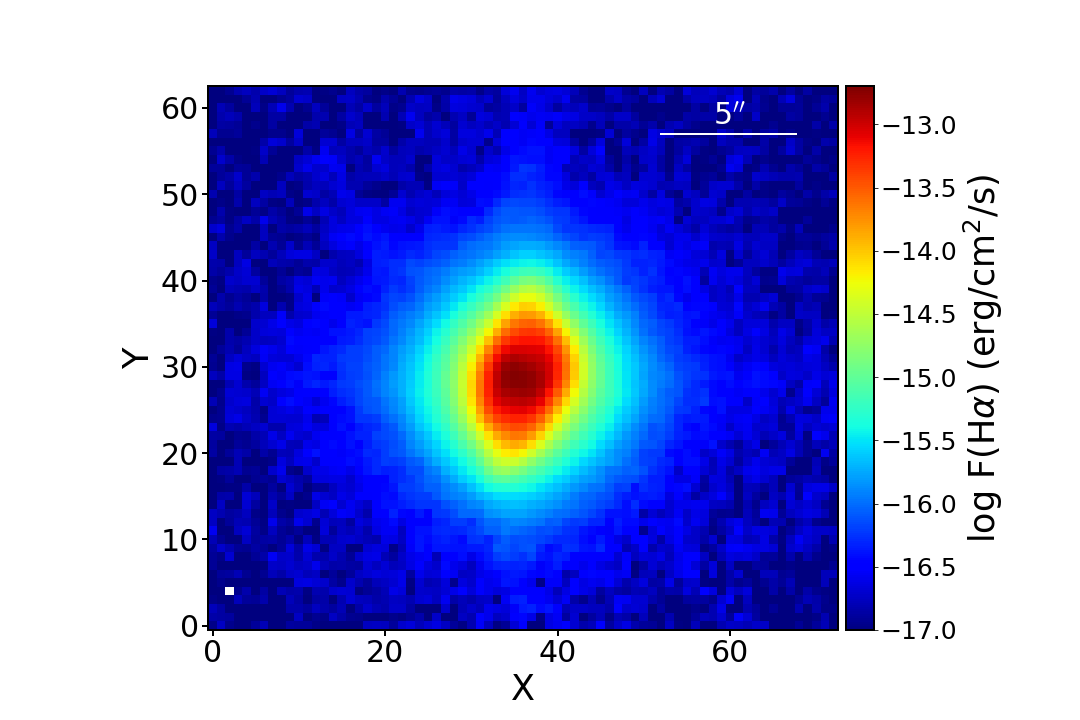}
		\includegraphics[width=\columnwidth]{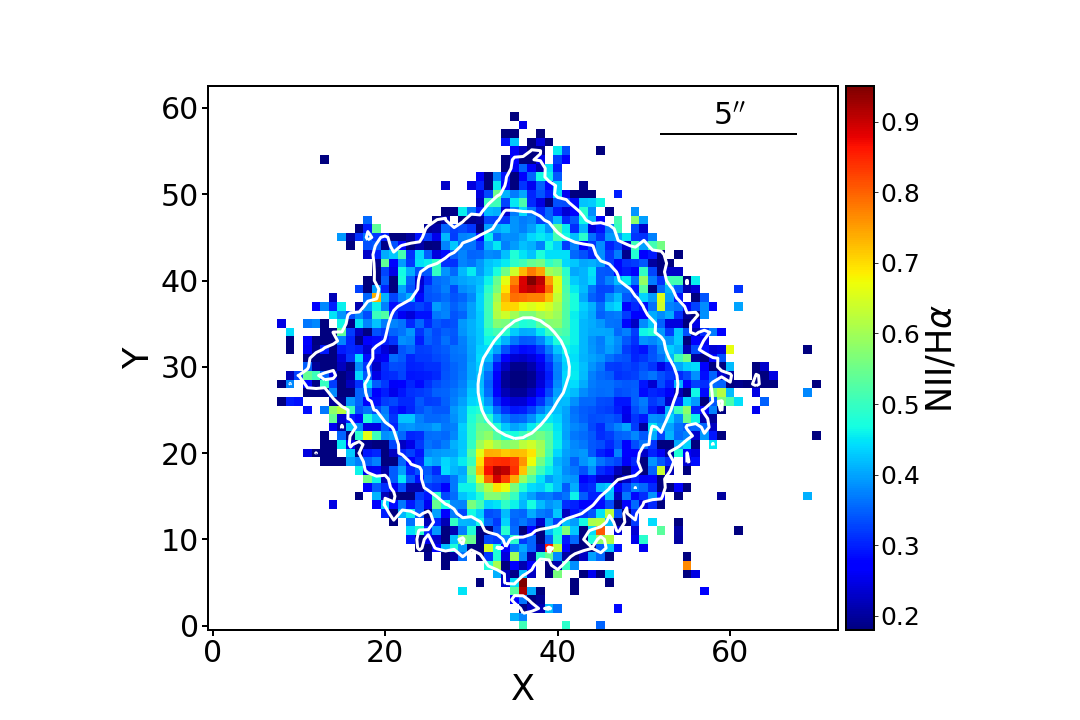}
		\includegraphics[width=\columnwidth]{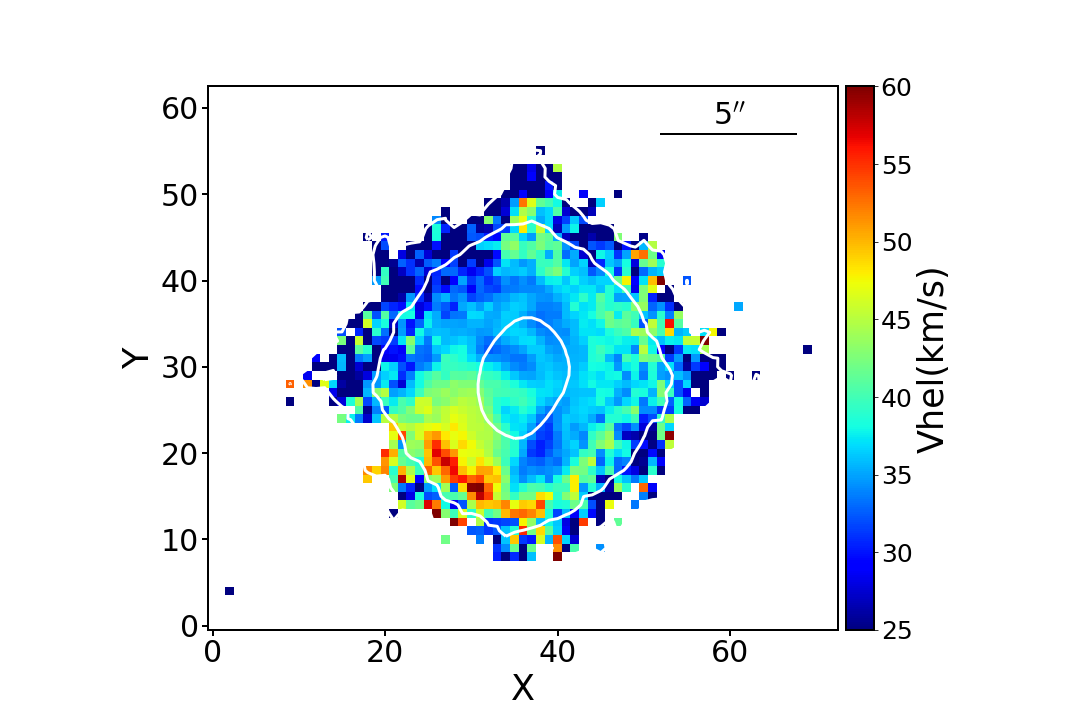}
		\includegraphics[width=\columnwidth]{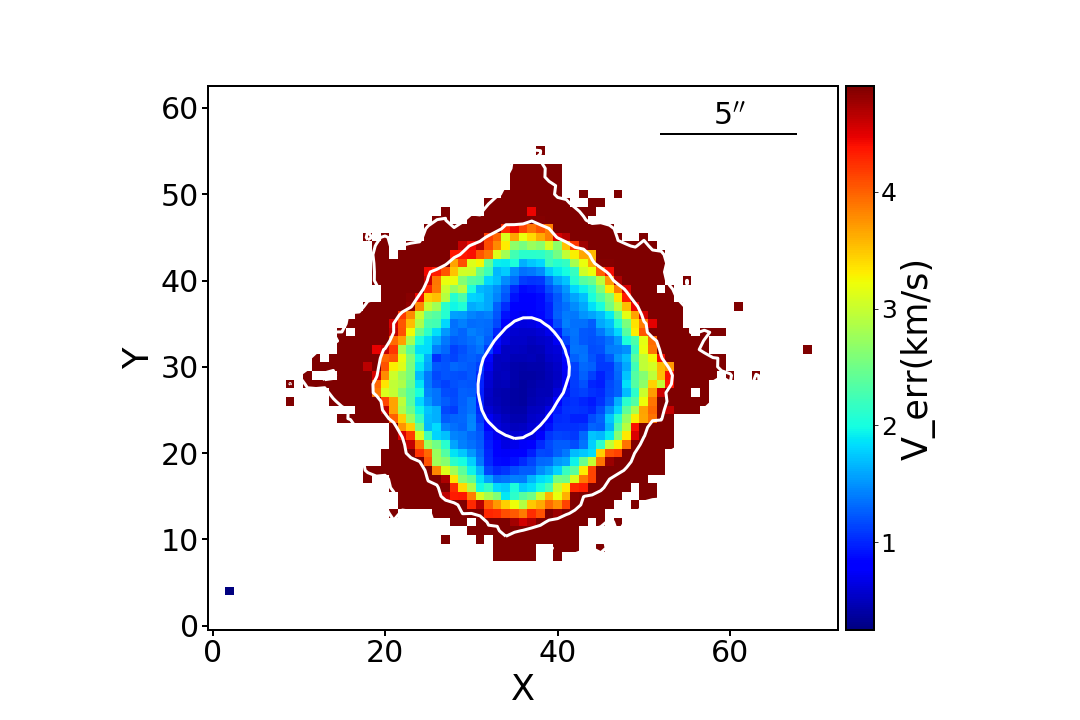}
    \caption{Planetary nebula M1-71 as seen from SITELLE. \textit{Upper panel}: H$\alpha$ (left) and [N\II]\,$\lambda$6584/H$\alpha$ ratio (right) images. \textit{Lower panel}: Heliocentric velocity map and formal uncertainty of the velocity, per pixel. Images are $22'' \times 20''$ with North at the top and East to the left.}
    \label{fig:M1-71a}
\end{figure*}

Figure~\ref{fig:M1-71a} shows the H$\alpha$ intensity and [N\II]\,$\lambda$6584/H$\alpha$ maps of the nebula, as well as velocity and velocity 
error maps derived with ORCS using all the lines. 
A clear bipolarity is obvious in the latter image, with line ratios varying between 0.15 in the center and 0.95 in the two ``caps''. The fits with a sincgauss function using ORCS are barely distinguishable from the data over two orders of magnitude in flux.
 
\begin{figure}
		\includegraphics[width=\columnwidth]{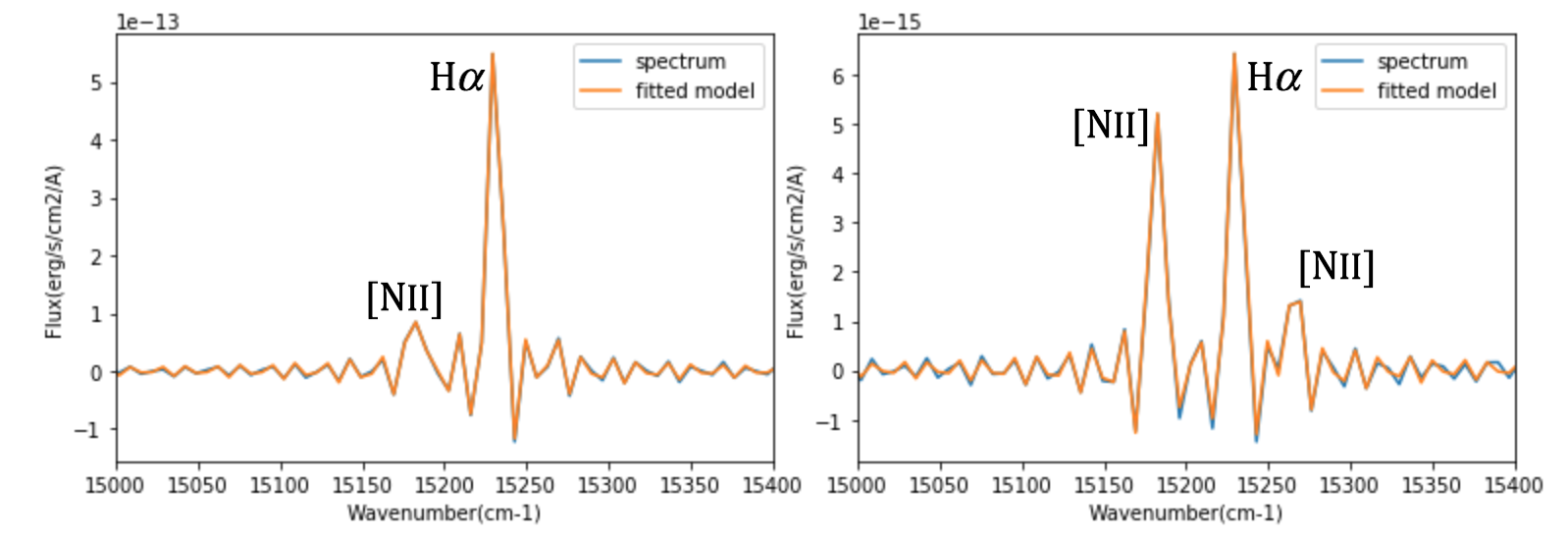}
    \caption{Spectra of 1\,arcsec$^2$ extracted in the center of M71 (left) and in the upper [N\II]-rich cap (right). Fits (orange) with sincgauss functions using ORCS are superimposed on the real spectrum (blue); the agreement is remarquable, over two orders of magnitude in flux. An identical fitting procedure was used on individual pixels to produce the maps shown in Figure~\ref{fig:M1-71a}. }
    \label{fig:M1-71b}
\end{figure}

The extracted radial velocities and their uncertainties are presented in the lower panel of Figure~\ref{fig:M1-71a}. The radial velocity map displays a very complex pattern. Note here that this map was obtained fitting a single component to the data, as the spectral resolution is clearly insufficient to detect line splitting. We have attempted to fit two components, using a range of initial velocities, but the results were not convincing. However, we note that the arc seen at the southern edge of the central part, with velocities $\sim$\,50\,-\,60\,km\,s$^{-1}$, is clearly detected in the datacube at higher velocities. We measure an average velocity, integrated over the entire nebula, of +38.5\,km\,s$^{-1}$ (with an uncertainty of the order of 1\,km\,s$^{-1}$; see the lower right panel of Figure~\ref{fig:M1-71a}), in good agreement the value (+42\,$\pm$\,8\,km\,s$^{-1}$) measured by  \citet{2005A&A...436..967W}. Figure~\ref{fig:M1-71b} shows spectra (integrated over 9~pixels) extracted in the center of the nebula and in the brightest region of the upper
cap.

\begin{figure}
	\includegraphics[width=\columnwidth]{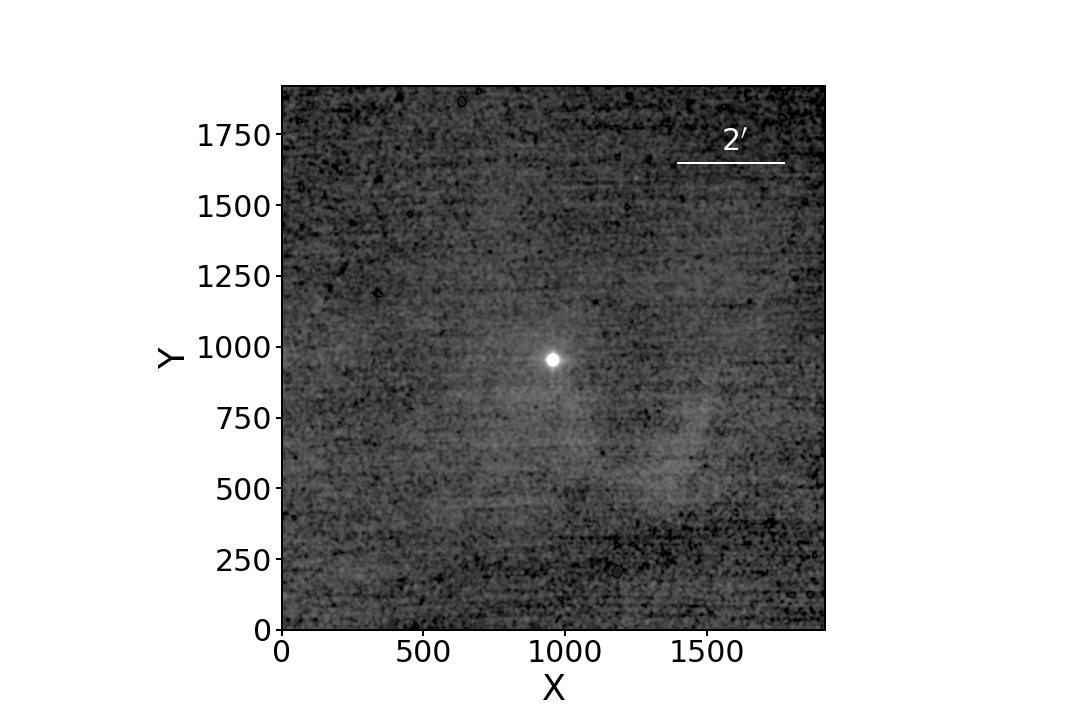}
    \caption{Integrated H$\alpha$ image of the entire SITELLE field ($11' \times 11'$) around M1-71, showing the faint, diffuse nebula. The horizontal stripes are artefacts from the CCDs. Stars have been removed and the image has been convolved with a 1.5$"$ gaussian kernel to increase the signal.}
    \label{fig:M1-71-nebul}
\end{figure}

Finally, we have detected a very low intensity, diffuse and morphologically complex H$\alpha$ structure spanning the entire field of view (Figure~\ref{fig:M1-71-nebul}). We cannot say with the current data if this structure is physically associated with M1-71 or merely an unrelated ionized cloud in the line of sight.

\subsection{Absorption Features in Galaxies}

\begin{figure*}
	\includegraphics[width=18cm]{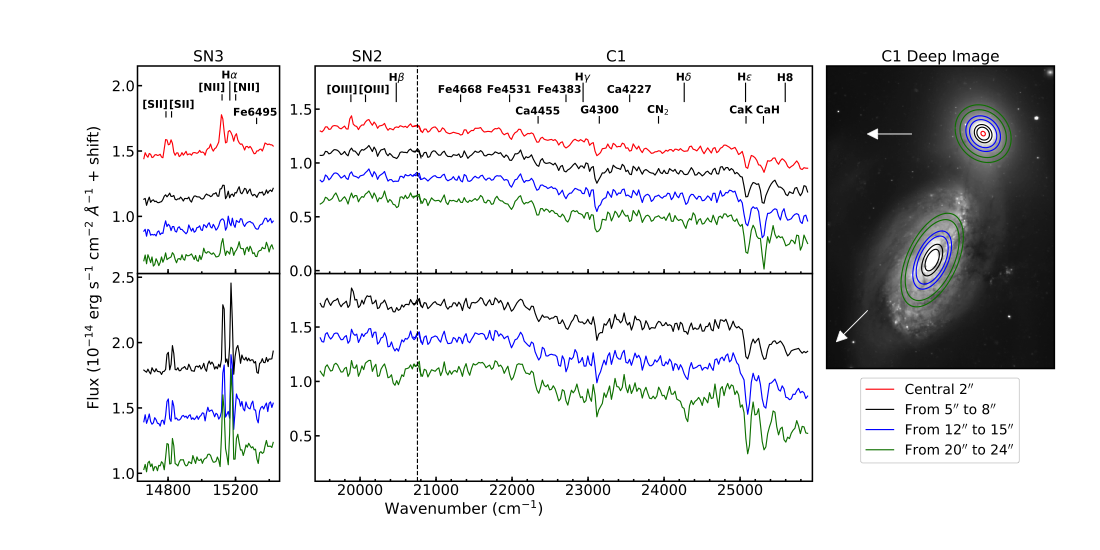}
    \caption{Spectra of annular regions in the galaxy members of Arp\,94. A portion of the C1 deep image is presented to the right, showing the position of the selected annuli for NGC\,3226 (the elliptical) and NGC\,3227 (the spiral). The de-projected internal and external radius of each annulus is given in the box bellow the image (using the same color code as for the spectra). The top and bottom panels of the left side of the image show the spectra for NGC\,3226 and NGC\,3227, respectively. The three filters used are identified at the top: SN3 to the left with SN2 and C1 in the middle. The dotted line indicates the overlapping wavelength region between SN2 and C1.  A small shift was applied to the flux in the different annuli for clarity of the plots: from the central to the more distant annulus, the shifts are for the elliptical: 0.85, 0.55, 0.25, and 0.0\,$\times$\,10$^{-14}$; and for the spiral: 1, 0.5, and 0.0\,$\times$\,10$^{-14}$. Emission lines from the ionized gas and absorption lines from the stellar populations are identified in the top panels.}
    \label{fig:arp94}
\end{figure*}

Contrary to a dispersive spectrograph, for which the photon noise at a given wavelength only depends on the flux from the source and its background at that particular wavelength, iFTS photon noise at each wavelength comes from the entire bandpass observed. Indeed, all photons from the bandpass are acquired at each mirror step. The use of a filter to shorten the bandpass width helps in reducing the noise, but an iFTS still remains much more sensitive to continuum photon noise than dispersive IFUs, and therefore is not optimal to study the absorption profiles from stellar population in low surface brightness galaxies. Nevertheless, our experience with SpIOMM \citep{2014AdAst2014E...9D} has shown that the central regions of elliptical galaxies can be studied with such an instrument. As a test for SITELLE, we have therefore targeted the interacting pair Arp\,94, composed of the SAB(s)a peculiar/Sy2  galaxy 
NGC\,3227 and its elliptical companion NGC\,3226, with filters C1, SN2, and SN3 in order to assess the instrument's capabilities in terms of absorption features for various surface brightnesses as well as to characterize the gaseous component of the pair. Figure~\ref{fig:arp94} shows the integrated spectra of a series of annuli centered on  nuclei of both galaxies in Arp\,94.

To obtain the spectra presented in Figure~\ref{fig:arp94}, the datacubes collected on different nights were first aligned using the bright stars present in the FOV. The sky background was subtracted using a median spectrum extracted from regions (the same for each datacube; using about 312\,000\,pixels) selected away from the galaxy pair. 

As shown on the deep image in Figure~\ref{fig:arp94}, annuli used to create the different galaxy spectra take into account the orientation of the objects in the sky. Values of the ellipticity, inclination, and position angle for the two galaxies have been estimated using the routine {\tt ellipse} from IRAF : 
i\,=\,60$^\circ$ and PA\,=\,155$^\circ$ for the spiral and $\varepsilon$\,=\,0.16 and PA\,=\,40$^\circ$ for the elliptical (these number are in good agreement with those from NED). For the purpose of  the demonstration here, no correction for the extinction or for the galaxy's internal motions has been applied prior to the summation of the spectra in each annulus. The choice of the position of the annuli was rather arbitrary, but their width was selected to display a similar flux (SNR\,$\simeq$\,7; as specified in the figure caption the spectra have been shifted by a small value to ease their comparison). A spectrum in the central region of the spiral galaxy is not shown, not like for the elliptical, as its known Seyfert signatures are strong and mitigate the clarity of the plots. A part of the SN2 and C1 wavelength range overlaps (near the dotted line in Figure~\ref{fig:arp94}). The flux agreement between these two filters indicates an uncertainty in the flux calibration of 10\% (and the C1 spectra have been shifted for a perfect match with the SN2 spectra in Figure~\ref{fig:arp94}).

In Figure~\ref{fig:arp94}, many of the Lick indicators \citep{1994ApJS...94..687W} used to characterize the stellar populations have been identified along with the strong emission lines generally used to study the ionized gas. Many of the unidentified structures in the spectra are more absorption lines or weak emission lines, also shaped by the instrument sinc profile. Clearly the C1 filter can be useful to study absorption features in different positions in these galaxies. In the SN3 filter, the stellar absorption feature Fe6495 is seen, displaying a different shape from the elliptical to the spiral galaxy. As expected, the LINER nature of the elliptical galaxy is responsible for strong and broad emission lines in the SN3 filter, but it is interesting to see here the presence of these emission lines further away from the galaxy center (for example, the [O\III]\,$\lambda$5007 emission is still observed in the annulus from 21$'$ to 26$'$). In the case of the spiral galaxy, the SN3 emission lines seen through the disk are the signature of H\II\ regions (many easily seen in the C1 deep image) but also of a diffused ionized gas component (as suggested in some cases by the high [N\II]/H$\alpha$ line ratio). A pure absorption component for 
the H$\alpha$ line is not seen in these data while an H$\beta$ absorption line can be detected at larger radii (although probably contaminated by an emission component in many cases). 

\subsection{Emission-line Galaxies in Clusters and Serendipity}

Nearby galaxy clusters in the z\,=\,0.04\,-\,0.23 range have been targeted, aiming at detecting the [O\II] line in the filters C1 (Abell\,1413, Abell\,2261) and SN2 (HETDEX Pilot Survey - COSMOS field), as well as the H$\alpha$ line in the filters SN3 (Abell\,168) and C4 (Abell\,2390). Moreover, background emission-line galaxies are always detected in all datacubes aiming away from the Galactic plane. Detailed analyses of these data will be published elsewhere, but we present here some representative results.

\subsubsection{Abell\,168}
\begin{figure*}
\includegraphics[width=\columnwidth]{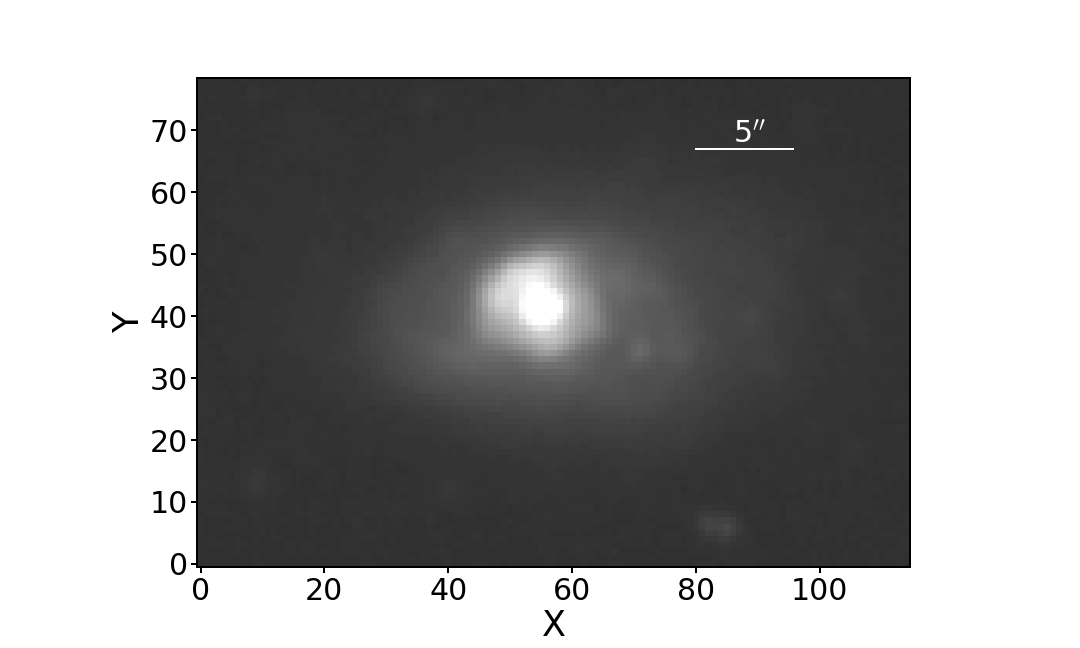}
\includegraphics[width=\columnwidth]{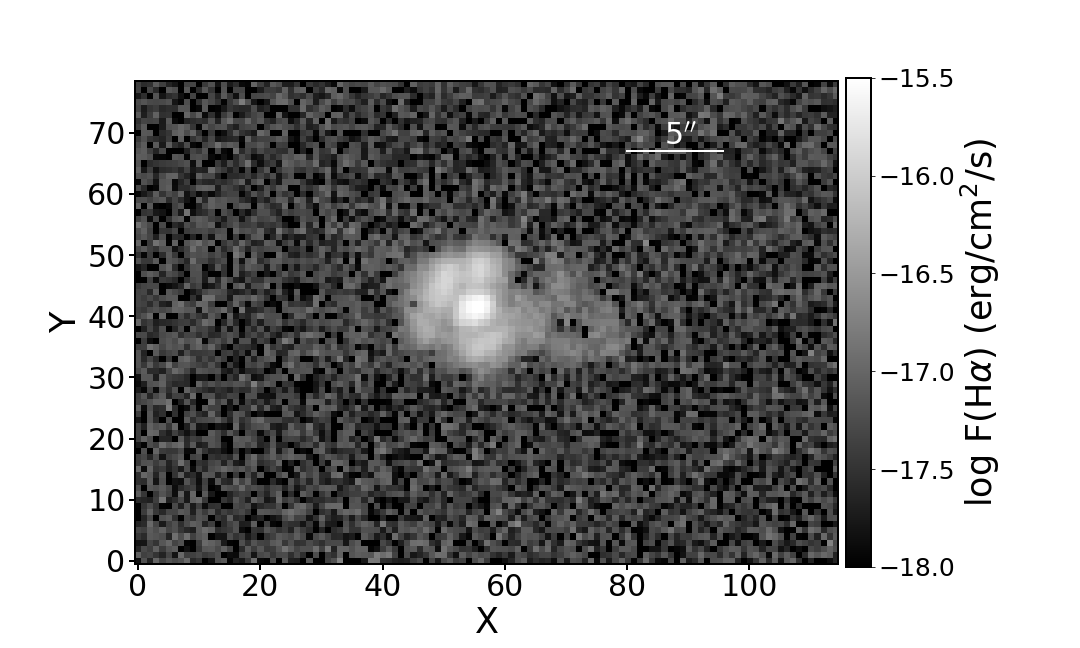}
\includegraphics[width=\columnwidth]{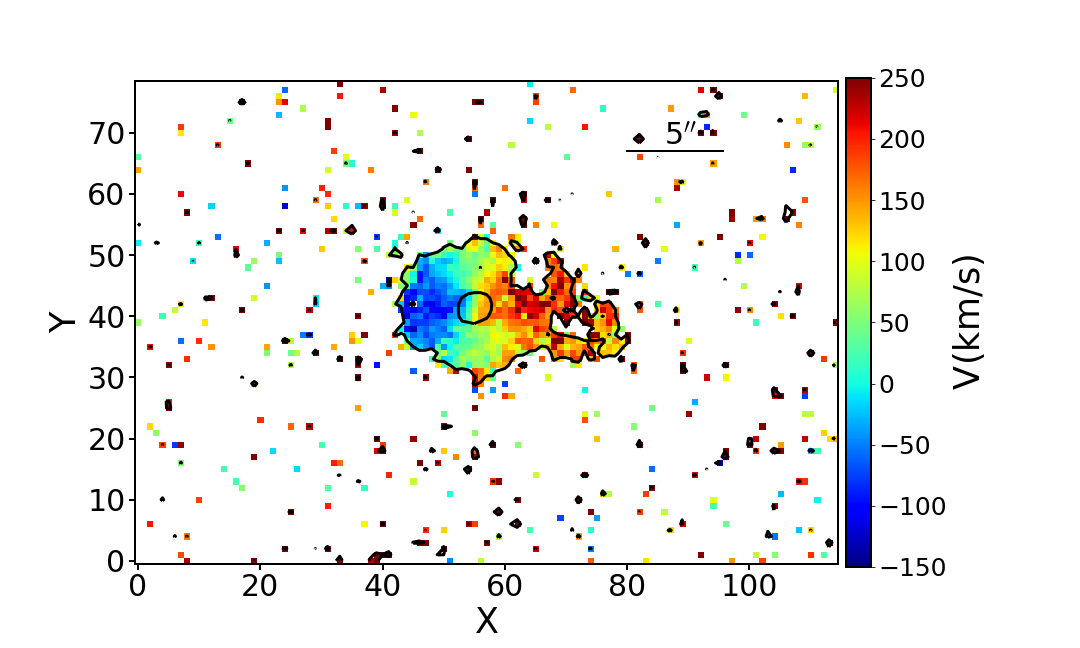}
\includegraphics[width=\columnwidth]{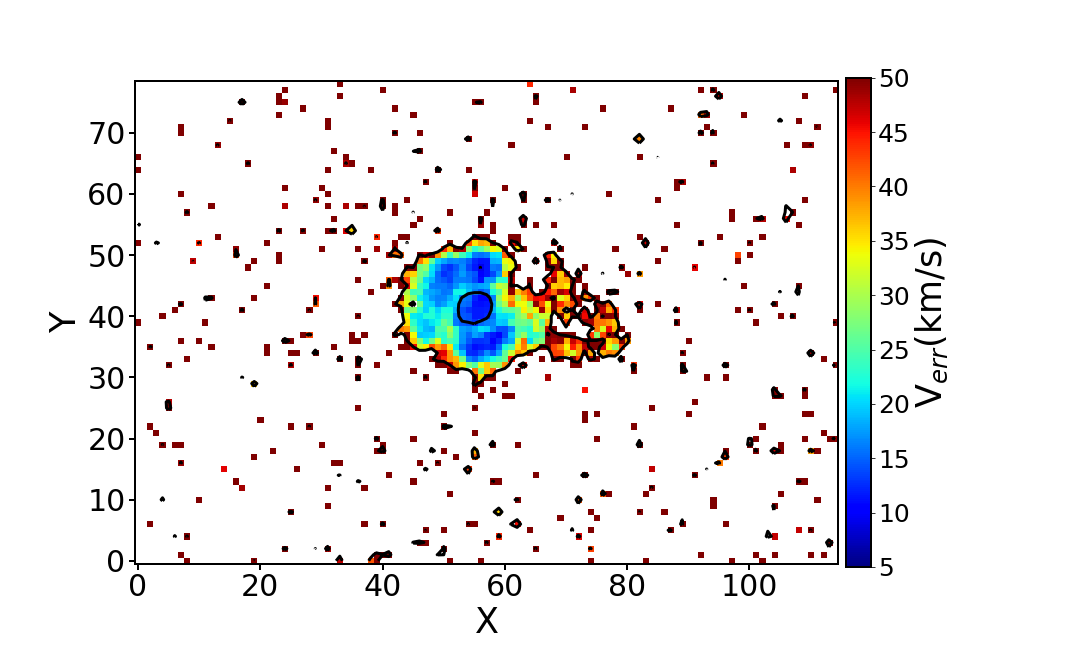}
    \caption{SDSS J011508.22+001337.5, a spiral galaxy in the Abell\,168 cluster, as seen from a SITELLE SN3 datacube. The upper panel shows the deep SN3 image (left) and the H$\alpha$ emission detected by ORCS. The lower panel shows the velocity field (left), relative to the systemic heliocentric velocity of 12850\,km\,s$^{-1}$, as well as the formal uncertainty on the velocity, per pixel (right). Each plot is $37''  \times 25''$ on a side, with East to the left and North at the top.}
    \label{fig:A168}
\end{figure*}

The average redshift of Abell\,168, z\,=\,0.045, places the H$\alpha$ line at the red edge of the SN3 filter and superimposed on a series of bright night-sky OH lines. Nevertheless, all members known to display emission lines in their spectra within the filter bandpass were also detected in the SITELLE datacube, allowing a 2D mapping of the emission. The most interesting case is that of SDSS J011508.22+001337.5, shown in Figure~\ref{fig:A168}. While the deep image shows a distorted spiral galaxy, the H$\alpha$ image reveals a bright core surrounded by a ring of very active star-forming regions, as well as more diffuse H$\alpha$ emission in the inner western arm. The H$\alpha$ flux from the core, within a 3\arcsec\ diameter aperture (corresponding to the SDSS fiber) determined by ORCS is 8.55\,$\times$\,10$^{-15}$\,erg~s$^{-1}$\,cm$^{-2}$, in excellent agreement with that provided by SDSS. We measure a total 
H$\alpha$ flux from the galaxy of 2.25\,$\times$\,10$^{-14}$\,erg~s$^{-1}$\,cm$^{-2}$. The lower panel of Figure~\ref{fig:A168} shows the velocity field of the ionized gas in this galaxy. 

\subsubsection{HETDEX Pilot Survey - COSMOS Field}

\begin{figure*}
	\includegraphics[width=5.5cm]{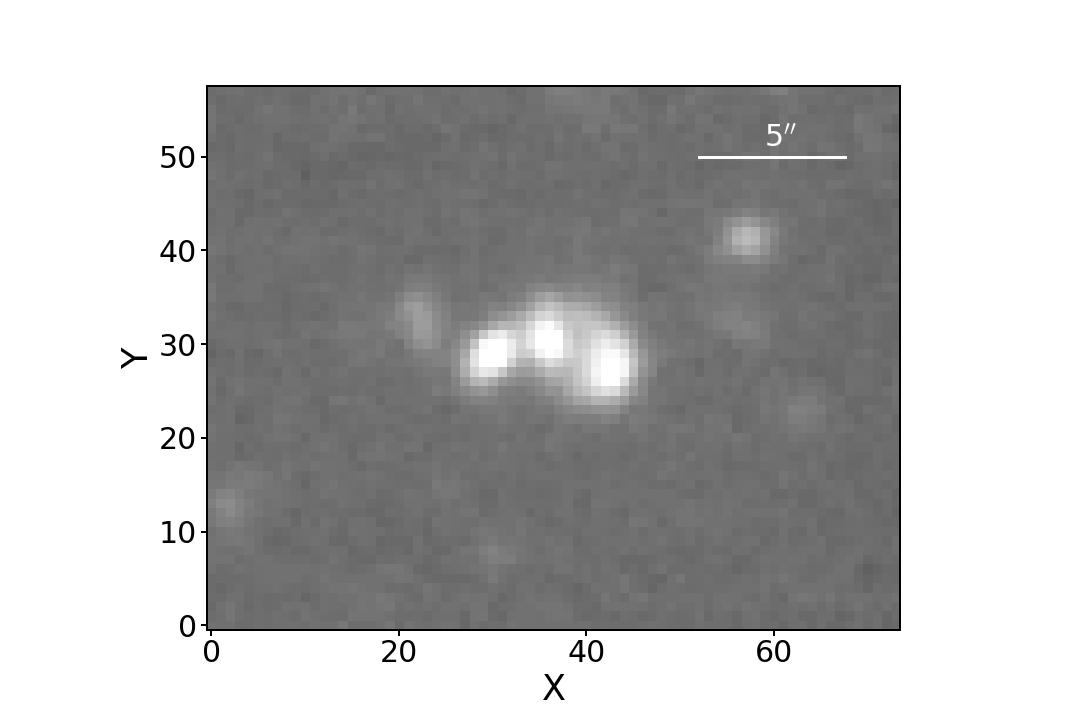}
		\includegraphics[width=5.5cm]{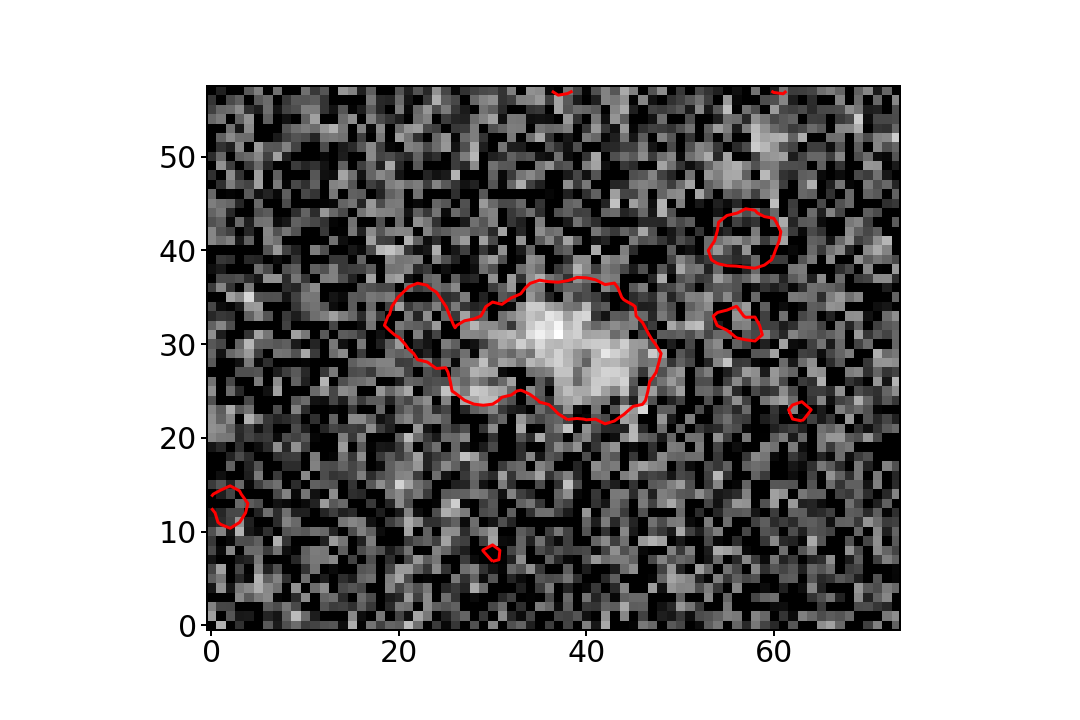}
		\includegraphics[width=5.0cm]{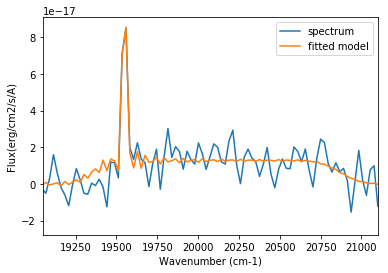}
    \caption{Continuum (left) and emission-line (middle) images, and the integrated spectrum (right) of HPS248. 
    An ORCS fits to the emission line ([O\II] at z\,=\,0.372) is superimposed. }
    \label{fig:HPS1}
\end{figure*}

\begin{figure*}
\includegraphics[width=6.0cm]{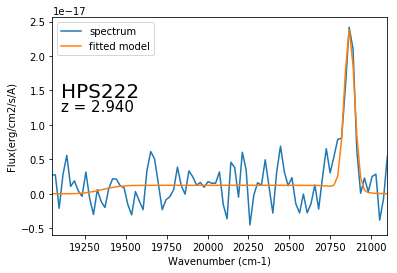}
\includegraphics[width=6.0cm]{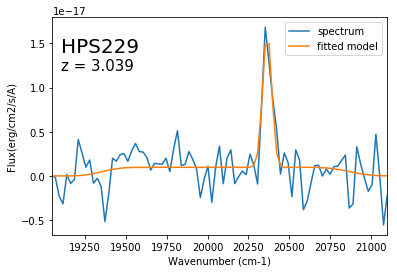}
\includegraphics[width=6.0cm]{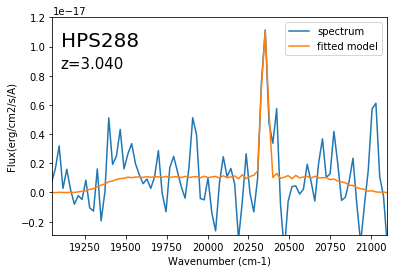}
\includegraphics[width=6.0cm]{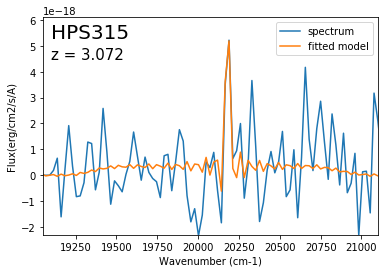}
    \caption{Spectra (blue) of four Ly-$\alpha$ emitters detected in the HPS COSMOS field, with ORCS fits (orange) superimposed. 
The redshift is deduced from the fit. Notice how the sinc sidelobes, obvious in the fits of the narrow line of HPS315 and barely visible in the slightly broader fit to HPS288, disappear completely for the broad lines of HPS222 and HPS229. }
    \label{fig:HPS2}
\end{figure*}

A field centered at 10:00:04, +02$^\circ$16$'$04$''$, which was included in the HETDEX Pilot Survey (HPS; \citealt{2011ApJS..192....5A}) was observed with the SN2 filter. This datacube was however plagued with some stray light probably originating from the metrology laser, resulting in bright spots scattered in the upper part of the field. Forty-one emission-line sources with a line falling within the SN2 filter bandwidth were known in this area, providing a good testbed to assess SITELLE's detectability limits. 29 sources were detected (success rate of 71\%), including 25 out of 31 (81\%) with an emission-line flux above 10$^{-16}$\,erg~s$^{-1}$\,cm$^{-2}$ and 4 out of 10 (40\%) below this limit. Figure~\ref{fig:HPS1} shows the continuum and emission-line images, as well as the integrated spectrum of HPS248, which is very likely a group of 3, perhaps 4 interacting galaxies at a redshift of z\,=\,0.372. Four out of the seven known Ly-$\alpha$ emitters within the redshift range covered by the SN2 filter have been recovered by SITELLE; their spectra are displayed in  Figure~\ref{fig:HPS2}.
Among them, HPS315 shows one of the weakest emission lines among our detections in this field, with an integrated emission-line flux of 
8.5\,$\times$\,10$^{-17}$\,erg~s$^{-1}$\,cm$^{-2}$. It is also the most distant source yet detected with SITELLE. We did not detect the Ly-$\alpha$ emitters HPS184, HPS 253 nor HPS 310.

\subsubsection{Background Galaxies in the Abell\,2390 Field}

\begin{figure}
	\includegraphics[width=\columnwidth]{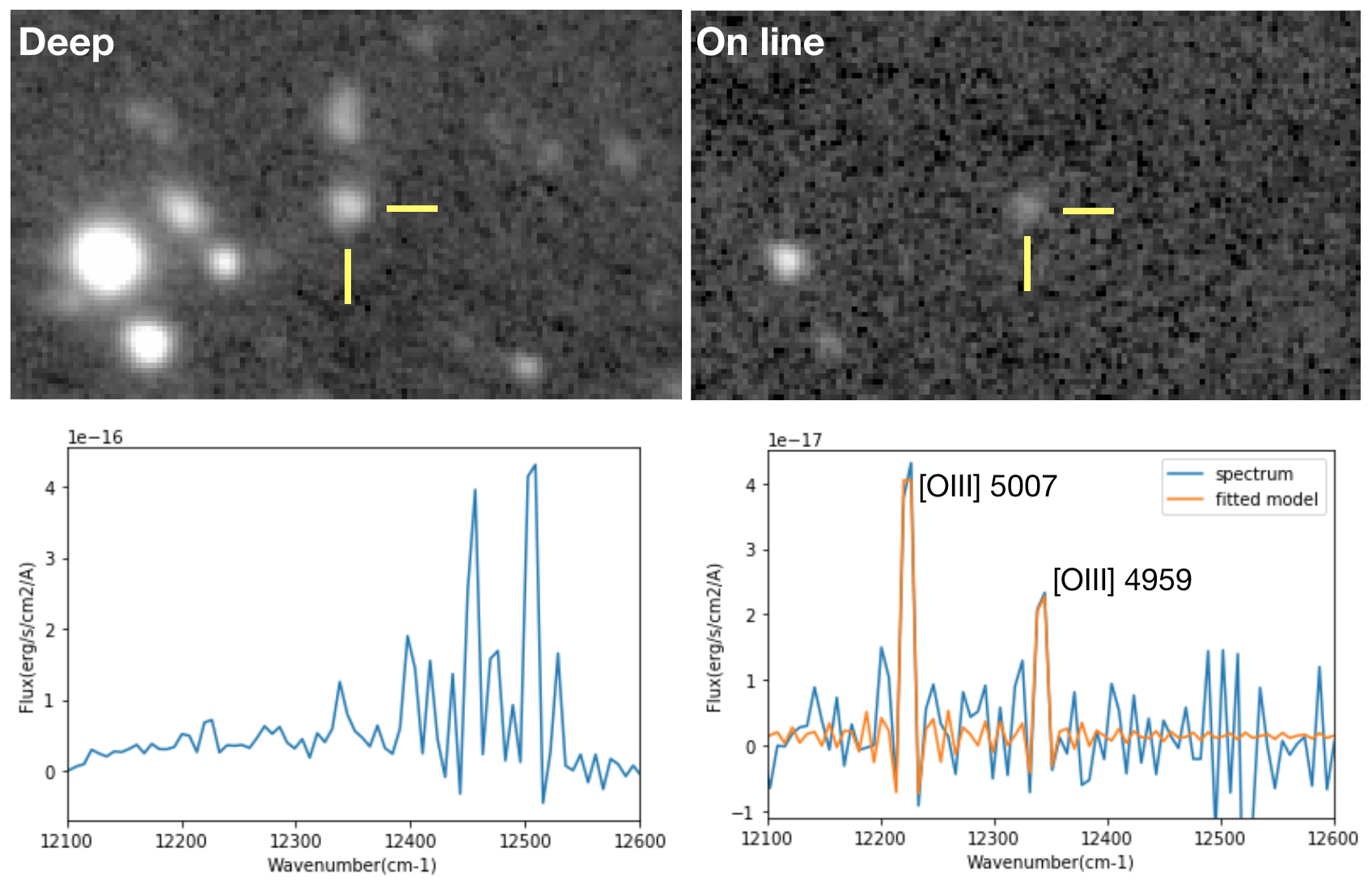}
    \caption{Emission-line galaxy in the field of Abell\,2390. The upper panel shows $41'' \times 24''$ deep and on-line images from the datacube,
    centered at 21h53m41s, +17$^\circ$29$'$52$''$. The lower panel shows the raw (left) and sky-subtracted integrated spectra (right) of the galaxy, as well as the ORCS fit to the [O\III]\,$\lambda\lambda$4959,5007 doublet. The redshift of the galaxy is 0.634. }
        \label{fig:A2390}
\end{figure}

The C4 filter was specially designed to detect H$\alpha$ emitters in the narrow redshift range of z\,=\,0.21\,-\,0.25, corresponding to a narrow window in the night-sky OH forest. Some sky lines are nevertheless present, and the continuum is stronger than in the other SITELLE bandpasses, contributing to the photon noise and therefore reducing the detectability of emission-line sources. But it is also in this range that SITELLE is the most efficient: the modulation efficiency, optical transmission and CCD quantum efficiency are at their best around 800\,nm. The first cluster to be observed using the C4 filter was 
Abell\,2390 (PI: Howard Yee), where more than 100 emission-line members have been detected (Yee et al., in preparation). Interestingly, the 
[O\III]\,$\lambda$5007 lines falls within this filter for redshifts z\,$\simeq$\,0.59\,-\,0.65 and a few of these outliers have been detected. 
Figure~\ref{fig:A2390} shows continuum and on-line images, as well as the spectrum of one of them to illustrate SITELLE's capability to accurately subtract the strong sky background ; we have independently fitted the two lines from the doublet and their velocities came out within 5\,km\,s$^{-1}$ of each other, 
at a redshift of 0.634. The [O\III]\,$\lambda$5007 flux of this source is 9.1\,$\pm$\,1.1\,$\times$\,10$^{-17}$\,erg~s$^{-1}$\,cm$^{-2}$.

\section{Conclusions}
In this paper, we have demonstrated that SITELLE is a very versatile instrument, capable of providing spatially resolved spectra of a variety of targets, from Galactic planetary nebulae to nearby galaxies and more distant galaxy clusters, in selected wavebands across the visible spectral range.

SITELLE's unique advantages for studying emission line objects are driving the CFHT Large Program SIGNALS (Star formation, Ionized 
Gas, and Nebular Abundances Legacy Survey). This project aims at observing a volume-limited sample of local, extended galaxies 
(D\,<\,10\,Mpc) with active massive star formation. Roughly 35 large nearby galaxies will be observed over 350 hours of observing time spread 
over 4 years at CFHT; observations have begun in October 2018. Each field will be observed with the SN1, SN2, and SN3 filters, with a 
resolution R\,=\,1000 for SN1 and SN2, and R\,=\,5000 for SN3. With an average spatial resolution of 20\,pc, this survey will provide the 
largest, most complete and homogeneous database of spectroscopically and spatially resolved extragalactic H\II\ regions ever assembled 
(Rousseau-Nepton et al., in preparation). The main goals are: 1) to quantify the impact of the surrounding environment on the star 
formation process; 2) to link feedback processes to the small-scale chemical enrichment and dynamics around star-forming regions; 
and 3) to measure variations of the resolved star formation rate with respect to indicators used for high redshift galaxy surveys. 
The SIGNALS dataset will be extremely rich. Notwithstanding the main study focusing on H\II\ regions, complementary results will also 
be obtained on supernova remnants, planetary nebulae, and background emission-line objects.

\section*{Acknowledgements}
Based  on  observations  obtained  with  SITELLE,  a  joint project  of  Universit\'e  Laval,  ABB,  Universit\'e  de  Montr\'eal,
and  the  Canada-France-Hawaii  Telescope  (CFHT)  which is  operated  by  the  National  Research  Council   
of Canada,  the  Institut  National  des  Sciences  de  l'Univers  of the  Centre  National  de  la  Recherche  Scientifique 
of France, and the University of Hawaii. The authors wish to recognize and acknowledge the very significant
cultural role that the summit of Mauna Kea has always had within
the indigenous Hawaiian community. We are most grateful to have
the opportunity to conduct observations from this mountain.
LD, CR, GJ and ST are grateful to the Natural Sciences and Engineering Research Council of
Canada, the Fonds de Recherche du Qu\'ebec, 
and the Canada Foundation for Innovation for funding. LD and CR are very grateful to the Department of Physics and Astronomy, University of Hawaii Hilo, as well as the CFHT, for having provided them with an ideal and very enjoyable working environment during their sabbatical stay on the Big Island.

\bibliographystyle{mnras}
\bibliography{biblio} 

\bsp	
\label{lastpage}
\end{document}